\documentclass[manusript]{aastex7}
\usepackage{stix2}
\usepackage[T1]{fontenc}
\usepackage[]{graphicx,tabularx,ctable,color,colortbl,amsmath,amsfonts,float,bm}
\usepackage{anyfontsize}

\shortauthors{}

\begin{document}

\title{Signatures of Large-Scale Magnetic Field Disturbances and Switchbacks in Interplanetary Type III Radio Bursts}

\author[0000-0003-1967-5078]{Daniel L. Clarkson}
\affiliation{School of Physics \& Astronomy, University of Glasgow, Glasgow, G12 8QQ, UK}
\email{daniel.clarkson@glasgow.ac.uk}

\author[0000-0002-8078-0902]{Eduard P. Kontar}
\affiliation{School of Physics \& Astronomy, University of Glasgow, Glasgow, G12 8QQ, UK}
\email{eduard.kontar@glasgow.ac.uk}

\begin{abstract}
Type III solar radio bursts are driven by non-thermal electron beams travelling along heliospheric magnetic fields, with the radio emission frequency drift-rate determined by the beam speed and the plasma density profile. Analysing beam kinematics inferred from the drift-rate reveals behaviour inconsistent with the emitter moving radially through smooth, monotonically decreasing density. We examine whether these features are driven by disturbances in the guiding magnetic field direction, such as switchbacks, rather than plasma inhomogeneities along the beam path. Using simulations and remote observations of 24 interplanetary type III bursts observed by Parker Solar Probe, we relate measured drift-rate variations to local field deflections. In 50\% of events, we identify disturbances above a $2\sigma$ noise level that can be attributed to perpendicular deflections of the field between (0.7--1.7)~R$_\odot$, over scales (1.8--6.4)~R$_\odot$ at heliocentric distances (9--30)~R$_\odot$. 
The features correspond to either density changes of (10--30)\%, or deflections of the field direction by (23--88)$\arcdeg$. Further, beam transport simulations show field direction perturbations produce additional observational signatures in type III bursts: delayed emission, intensity breaks, and enhanced emission resembling stria fine structures. In addition, we identified four bursts where the observed variations are more plausibly explained by field deflections, possibly in the form of magnetic switchbacks, than by unrealistically large density changes along the field line. The results show that variations in type III burst profiles can arise from magnetic as well as density fluctuations, and demonstrate the value of type III bursts as remote probes of inner-heliospheric structure at kilometric wavelengths.
\end{abstract}


\section{Introduction}

Solar radio bursts are intrinsically linked to the motion of the emitting source through the coronal and heliospheric plasma. The release of magnetic energy in flares accelerates electrons to non-thermal velocities 
\citep[e.g.][]{2011SSRv..159..107H} that stream through the heliosphere 
along pervading magnetic field lines 
\citep{2008A&ARv..16....1P, 2017LRSP...14....2B} 
that on large-scales form a Parker spiral \citep{1958ApJ...128..664P}. Electron transport is largely constrained along magnetic field lines, 
so kinetic models are often formulated in one dimension \citep[e.g.][]{1985srph.book..253G, 2001SoPh..202..131K, 2021NatAs...5..796R}.
The electron beams propagate at a substantial fraction of the speed of light often reaching around $c/3$ \citep[e.g.][]{1985srph.book..289S}, 
with speeds $\lesssim10\%$ of $c$ reported for interplanetary burst observations 
\citep[e.g.][]{1987A&A...173..366D, 2015A&A...580A.137K}. During propagation, these beams can generate radio emission via the plasma emission process \citep{1985ARA&A..23..169D, 2001SoPh..202..131K}, 
producing signals near the local plasma frequency $f_\mathrm{pe}$,
where $f_\mathrm{pe}\left(n(r)\right)\,[\mathrm{kHz}]\simeq8.93 \sqrt{n(r)\,\mathrm{[cm^{-3}]}}$ and/or its harmonic $2f_\mathrm{pe}$, 
where $n(r)$ is the local plasma density.

In dynamic spectra, this emission appears as so-called type III bursts that spans decades in frequency 
\citep[see, e.g.][as reviews]{1985SoPh..100..537L,2011SSRv..159..107H,2017LRSP...14....2B}. 
The time of maximum intensity is often used to trace the position 
of the electron beam in space 
\citep[e.g.][]{2015A&A...580A.137K,2018ApJ...867..158R, Azzollini_2025}.
In a smoothly decreasing density plasma, an electron beam moving along a radial path would produce a type III burst whose drift-rate gradually decreases over time. 
As a function of observing frequency, the drift-rate is approximately characterised by the statistical 
relation $\mathrm{d}f/\mathrm{d}t\,[\mathrm{MHz~s^{-1}}]=-0.01f^{1.84}$ \citep{1973SoPh...29..197A} covering (0.075--550)~MHz. 
Similarly, \cite{1999A&A...348..614M} find the relation $\mathrm{d}f/\mathrm{d}t\,[\mathrm{MHz~s^{-1}}]=-0.0074f^{1.76}$ between (0.04--140)~MHz, and \cite{2009POBeo..86..287V} find $\mathrm{d}f/\mathrm{d}t\,[\mathrm{MHz~s^{-1}}]=-0.02f^{1.8}$ between (0.125--3)~MHz.
The association of the observed burst frequency drift-rate with the exciter motion through the plasma naturally suggests 
that a reverse-drift (positive frequency drift-rate) implies 
that the exciter is propagating through an increasing density plasma. Consequently, radio bursts that exhibit a turnover in frequency, forming a J- or U-shaped pattern in dynamic spectra, 
are interpreted via an exciter moving along curved 
magnetic loop-like structures
\cite[e.g.][]{1958Natur.181...36M, 1974SoPh...39..451S,1992ApJ...391..380A,2017A&A...606A.141R,2021SoPh..296....1D, 2024ApJ...965..107Z}. 
Such features are therefore clear and obvious examples 
of large field variations visible in solar radio observations.

The frequency drift-rate can also infer the speed of the emitting source by relating the observed emission 
frequency over time $f(t)$ to a distance $r(t)$ 
with the assumption of a density structure $n(r)$ 
in the corona and heliosphere. 
This approach converts frequency drift-rate 
$\mathrm{d}f/\mathrm{d}t$ to velocity $\mathrm{d}r/\mathrm{d}t$ 
for fundamental or harmonic emission. 
In an inhomogeneous plasma, Langmuir wave refraction   
causes the average speed of interacting electrons 
to decrease with distance \citep{2001SoPh..202..131K}. 
As a result, exciter speeds are thought to decrease 
as they propagate through interplanetary space \citep{2015A&A...580A.137K, 2023SoPh..298...52L, Azzollini_2025}, contributing to a decrease in the drift-rate. A multi-spacecraft study by \cite{Azzollini_2025} shows that the primary driver of deceleration is the interaction of the beam-plasma structure with density inhomogeneities, 
with the type III exciter slowing with distance as $v(r)\propto r^{-0.3}$, 
within the uncertainty of that derived 
from observations \citep{2015A&A...580A.137K}.

Type III bursts also demonstrate fine structures, 
visible as multiple stria in the burst envelope. 
These so-called type IIIb bursts are often interpreted 
as the plasma emission process modulated by density 
fluctuations and were reproduced in a number of simulations \citep[e.g.][]{2001A&A...375..629K, 2021NatAs...5..796R}. 
Such fine structures were first observed in coronal radio emission due to the available temporal-spectral resolutions \citep[e.g.][]{1972A&A....20...55D, 2017NatCo...8.1515K}, and more recently reported 
in interplanetary bursts near 1~MHz \citep{2020ApJS..246...49P, 2023A&A...670A..20J} 
and as low as (30--80)~kHz \citep{2025ApJ...985L..27K}. 
Following the interpretation of stria generation in coronal bursts, interplanetary stria are also suggested to be due to density fluctuations.
Noteworthy, the time-frequency resolution of 
space-based observations is normally much lower
than ground-based telescopes that complicates fine structure analysis \citep{2024ApJ...974L..18V}.

At the same time, in-situ magnetic field measurements show that fluctuations are common throughout the solar wind, where most variations lie perpendicular to the field direction \citep{2013LRSP...10....2B, 2013SSRv..178..101A}. 
Measurements from Parker Solar Probe \citep[PSP;][]{2016SSRv..204....7F} have revealed the prevalence of folds in the magnetic field known 
as magnetic switchbacks 
\citep{2019Natur.576..237B, 2019Natur.576..228K} 
with characteristic lengths along their major axis of about 1~R$_\odot$ \citep{2021A&A...650A...1L}. 
Their origin remains uncertain, 
with suggestions linking them to remnants of near-Sun transient events \citep{2020ApJS..246...45H}. 
Additionally, large-scale loop-like magnetic structures in the solar wind, such as helmet streamers or pseudostreamers, can also influence particle transport through their magnetic fields \citep[e.g.][]{2022ApJ...934...55L}. 

Given the turbulent nature of the solar atmosphere, this study considers whether changes in type III radio burst drift-rates can be explained not only by density fluctuations as often done before \citep{1972A&A....20...55D,2018SoPh..293...26M}, 
but also by magnetic field deviations. We also explore how field deviations can produce several observational signatures in interplanetary radio bursts, including stria fine structures.
In section \ref{sec:dfdt}, we outline the influences on the burst drift-rate together with the observational approach to infer the extent of magnetic field changes from dynamic spectra data. We apply the method to simulations of field line disturbances in section \ref{sec:Bfluc_sim}, 
and to observational data in section \ref{sec:observations}. 
In section \ref{sec:typeIII_sim}, 
we explore additional features that field disturbances impart into type III burst profiles using numerical simulations, along with observational examples. Section \ref{sec:summary} summarises the results.

\section{Physical Influences on Type III Burst Drift-Rates}\label{sec:dfdt}
        
Observationally, the temporal evolution of the frequency of maximum emission intensity is described by the drift-rate 
$\mathrm{d}f/\mathrm{d}t$:
    \begin{equation}
        \frac{\mathrm{d}f}{\mathrm{d}t} = \frac{\mathrm{d}f}{\mathrm{d}r}\frac{\mathrm{d}r}{\mathrm{d}t},
    \end{equation}
where $f$ is the emission frequency, $r$ is the radial distance from the Sun. Since the emission frequency (funadmental or harmonic) is given by local density, density perturbations $n\pm\delta{n}$ along the electron beam path produce corresponding variations in the emission frequency $f\pm\delta{f}$ due to the gradient term $\mathrm{d}f/\mathrm{d}r$. Fine structures such 
as striae, embedded within the broadband burst signal, 
are now imaged and attributed to such density fluctuations \citep[e.g.][]{2017NatCo...8.1515K, 2018ApJ...856...73C, 2020ApJS..246...49P, 2025ApJ...985L..27K} 
through modulation of the plasma emission process \citep{2001A&A...375..629K, 2021NatAs...5..796R}.
    
\begin{figure}[htb!]
    \centering
    \includegraphics[width=0.35\textwidth]{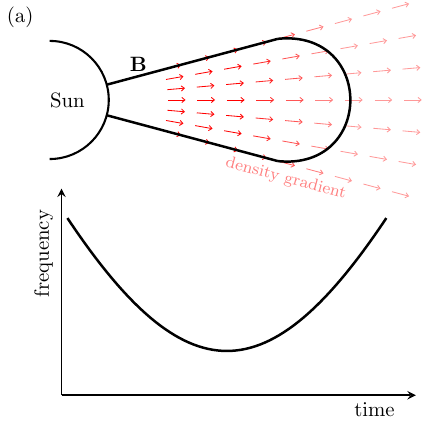}
    \includegraphics[width=0.35\textwidth]{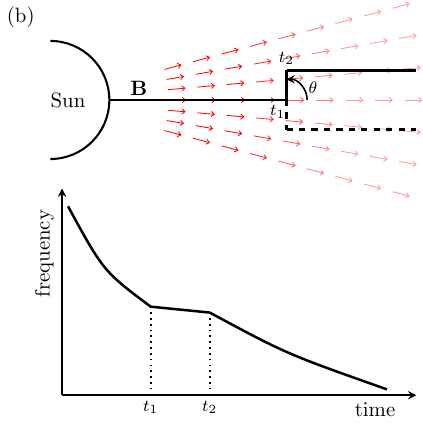}
    \caption{Cartoon showing the influence of a magnetic field disturbances on a type III radio burst profiles. In each column, a radio emitting source propagating along the magnetic structure would generate the radio burst profile displayed in the lower panels. \textit{(a)} Large-scale magnetic loop. \textit{(b)} A 90$\arcdeg$ deflection. In this case, a $\theta=\pm90\arcdeg$ deflection of the field in either direction would produce the same reduction in frequency drift-rate on the burst profile in dynamic spectra, shown in the lower panel. The red arrows show the radial density gradient with lighter color representing lower density.}
    \label{fig:cartoon}
\end{figure}

Alternatively, a disturbance in the magnetic field direction alters the position of the emitter with time and therefore affects the beam velocity direction, reducing the radial velocity $\mathrm{d}r/\mathrm{d}t$. 
Fluctuations in the radial velocity 
introduced by perturbations of the guiding magnetic field 
direction thereby modify the observed drift-rate,
mimicking a different density gradient.
Figure~\ref{fig:cartoon} illustrates two cases of magnetic field deflections: a large-scale magnetic loop, and a sharp $90\arcdeg$ field deflection, along with the corresponding 
reduction or reversal
of the type III burst drift-rate over the interval during which the emitter traverses the perturbed region.
Noteworthy, the opposite sign perturbation 
produces the \textit{same} spectral signature (Figure~\ref{fig:cartoon}).
The orientation of a field perturbation relative 
to the ambient density gradient controls 
how strongly it modifies the observed burst profile.
More complicated magnetic field perturbations 
are considered later in the paper. 

\subsection{Associating a Change in Drift-Rate with a Field Disturbance}\label{sec:method}

To relate a change in the burst drift-rate to a change in magnetic field direction with respect to a chosen reference, one can consider a two dimensional 
magnetic field $\mathbf{B}$ with radial and perpendicular components, so that $B^2 = B_r^2 + B_\perp^2$. 
For a magnetic field line satisfying     \begin{equation}\label{eq:field_line}
        \frac{\mathrm{d}r}{B_r} = \frac{\mathrm{d}r_\perp}{B_\perp},
    \end{equation}    
one can find the angle of deviation from the radial direction as
\begin{equation}
        \tan\theta=\frac{\mathrm{d}r_\perp}{\mathrm{d}r} = \frac{B_\perp/B}{B_r/B}.
\end{equation}
Thus, Figure~\ref{fig:cartoon} demonstrates a $90\arcdeg$ deviation.

Since $B_r/B=\left[1 - B_\perp^2/B^2\right]^{1/2}$, 
the perpendicular field fluctuation ratio 
can be written as
\begin{equation}\label{eq:Bperp_B}
        \frac{B_\perp}{B} = \frac{\mathrm{d}r_\perp/\mathrm{d}r}{\sqrt{1 + (\mathrm{d}r_\perp/\mathrm{d
        r)^2}}},
\end{equation}
which requires only the parallel $\mathrm{d}r$ and perpendicular $\mathrm{d}r_\perp$ displacements. 
These are inferred observationally by mapping 
fluctuations in frequency-space $f(t)$ to distance $r(t)$. 
Here, $\theta$ represents the local angular deviation of the field from a reference direction. A large value of $r_\perp$ does not necessarily correspond to a high value of $B_\perp/B$, since the latter depends on the spatial scale across which the perturbation occurs---that is, 
for a given perpendicular field displacement, a large spatial scale implies a smaller angular deviation than a small spatial scale. Observationally, scatter in the measured burst intensities is also expected from instrumental limitations, which introduces uncertainties into the deduced value of $r_\perp$. 
We define a threshold below which fluctuations in $r_\perp$ are attributed to noise in Appendix \ref{appendix:noise},
in the context of spacecraft observations.
    
    \begin{figure}[b!]
        \centering
        \includegraphics[width=0.49\textwidth]{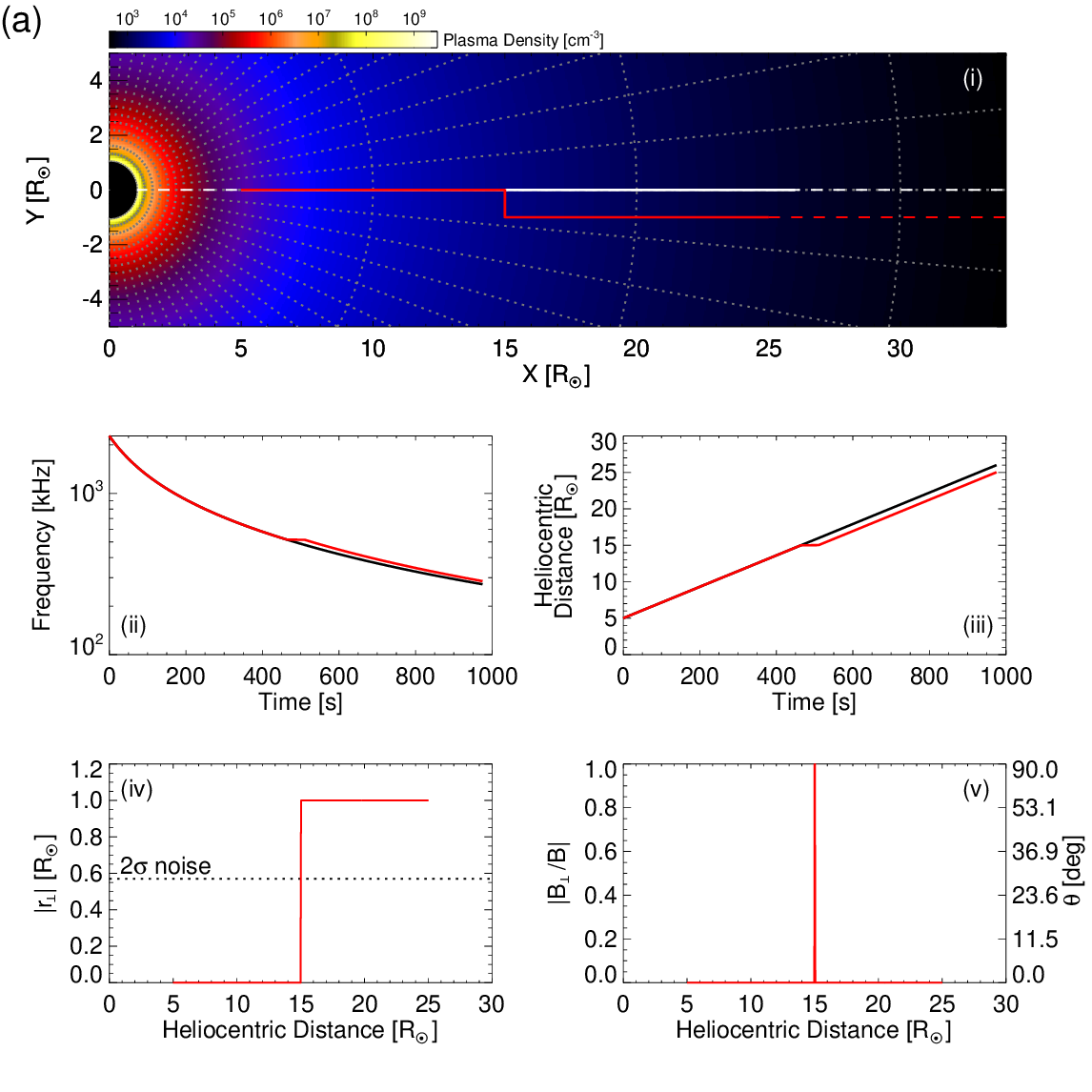}
        \includegraphics[width=0.49\textwidth]{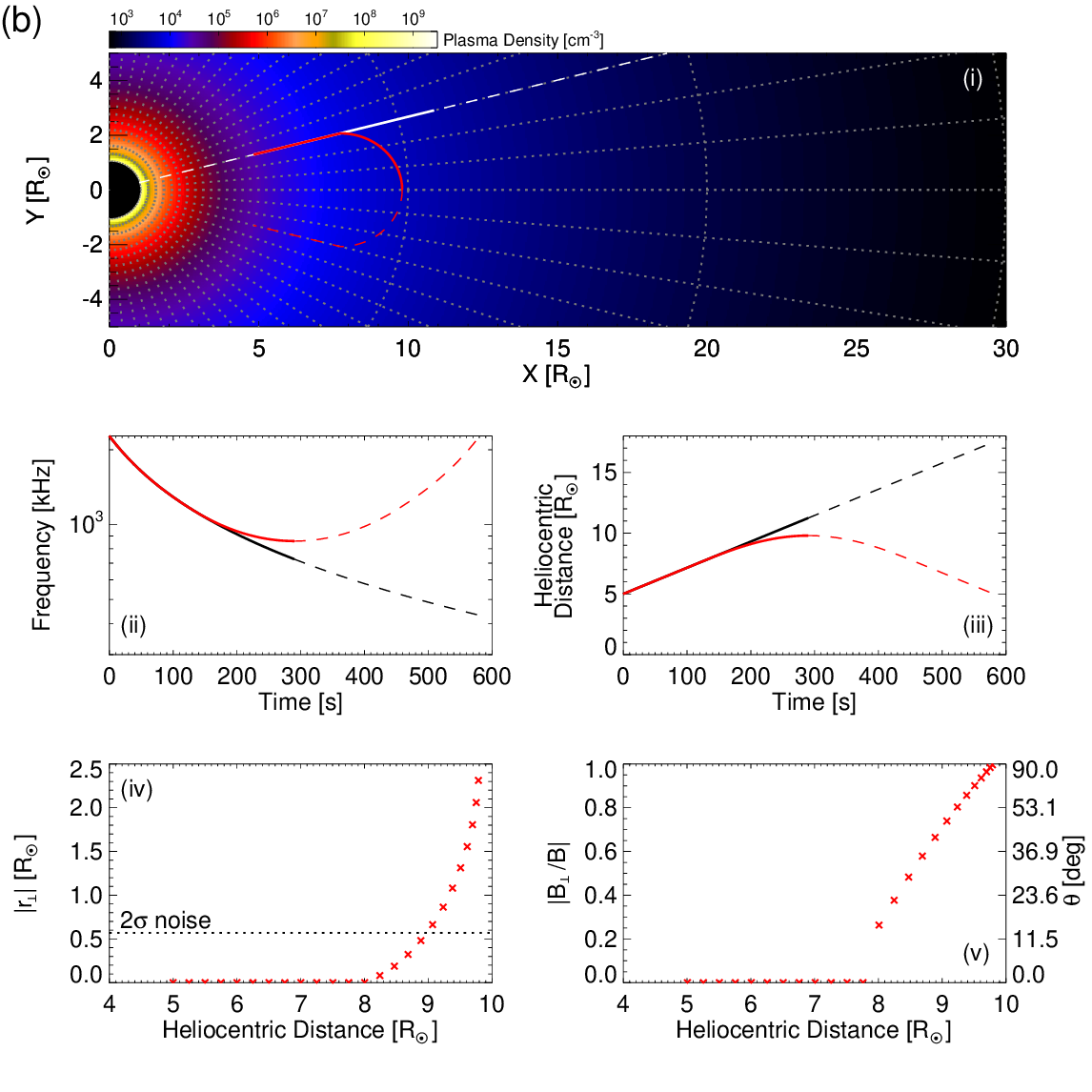}
        \caption{A $90\arcdeg$ field perturbation \textit{(a)} and a large-scale magnetic loop \textit{(b)}. 
        The electrons propagate along the radial and perturbed fields with an initial speed of $v_0=0.1c$ and decelerate as $v(r)\propto r^{-0.3}$. \textit{(i)} Paths of each magnetic field overlaid onto a 2D, radially symmetric heliospheric density map constructed using equation \ref{eq:Nparker}. The plasma density is shown by the colour gradient. The segments of each path analysed are shown as solid lines. The radial grid lines are separated by $5\arcdeg$. \textit{(ii)} The emission frequency experienced along each path. \textit{(iii)} The distance corresponding to each frequency from panels (ii). \textit{(iv)} Perpendicular deviation $r_\perp$ between the radial and curved paths. The dotted line shows the $2\sigma$ noise level determined in Appendix \ref{appendix:noise}. \textit{(v)} $B_\perp/B$ along the curved path.}
        \label{fig:kink_Jburst_sim}
    \end{figure}
    
\subsection{Magnetic Field Disturbances and Burst Profiles}\label{sec:Bfluc_sim}
        
To visualise field disturbances that could alter the burst profile, we first show two clear cases of field deflections: a $90\arcdeg$ deflection and a large-scale magnetic loop that leads to J- and U-bursts. We propagate an electron beam along each perturbed path (red) and along the radial lines (white), tracing the emission frequency $f_\mathrm{obs}$ experienced by the beam.
Following an observational approach, the distances along the paths are assigned a density using a spherically symmetric model \citep{2019ApJ...884..122K}:
    \begin{equation}\label{eq:Nparker}
        n(r)\,=4.8\times10^9\left(\frac{r}{R_\odot}\right)^{-14} + 3\times10^8\left(\frac{r}{R_\odot}\right)^{-6} + 1.4\times10^6\left(\frac{r}{R_\odot}\right)^{-2.3} \;\; [\mathrm{cm^{-3}}],
    \end{equation}
where $f_\mathrm{pe}\,\mathrm{[kHz]} = 8.93\sqrt{n(r)[\mathrm{cm^{-3}}]}$ is the plasma frequency. 
In Figure \ref{fig:kink_Jburst_sim}(a), 
the initial field segments are parallel, 
so $r_\perp$ and $B_\perp/B$ are zero. At 15~R$_\odot$, 
the field deflects by $90\arcdeg$, causing $r_\perp$ to increase to 1~R$_\odot$ and $B_\perp/B$ to jump to 1 (i.e. $\theta=90\arcdeg$). 
Beyond this point, the perturbed field line runs parallel to the unperturbed line, and $B_\perp/B$ returns to zero. 
    
Figure \ref{fig:kink_Jburst_sim}(b) shows a simple magnetic loop where the ascending and descending legs are taken to lie along the radial direction separated by $30\arcdeg$, with the loop section formed 
by a semi-circle with an apex near 10~R$_\odot$. 
Propagating an electron beam along the half-loop comprising the ascending leg up to the loop apex, 
and along the radial line, reveals that $r_\perp$ increases up to $2.5$~R$_\odot$ at the loop apex, 
well above the $2\sigma$ uncertainty level. 
Here, the field trajectory is fully perpendicular, 
with $B_\perp/B=1$.
    
    \begin{figure}[htb!]
        \centering
        \includegraphics[width=0.49\textwidth]{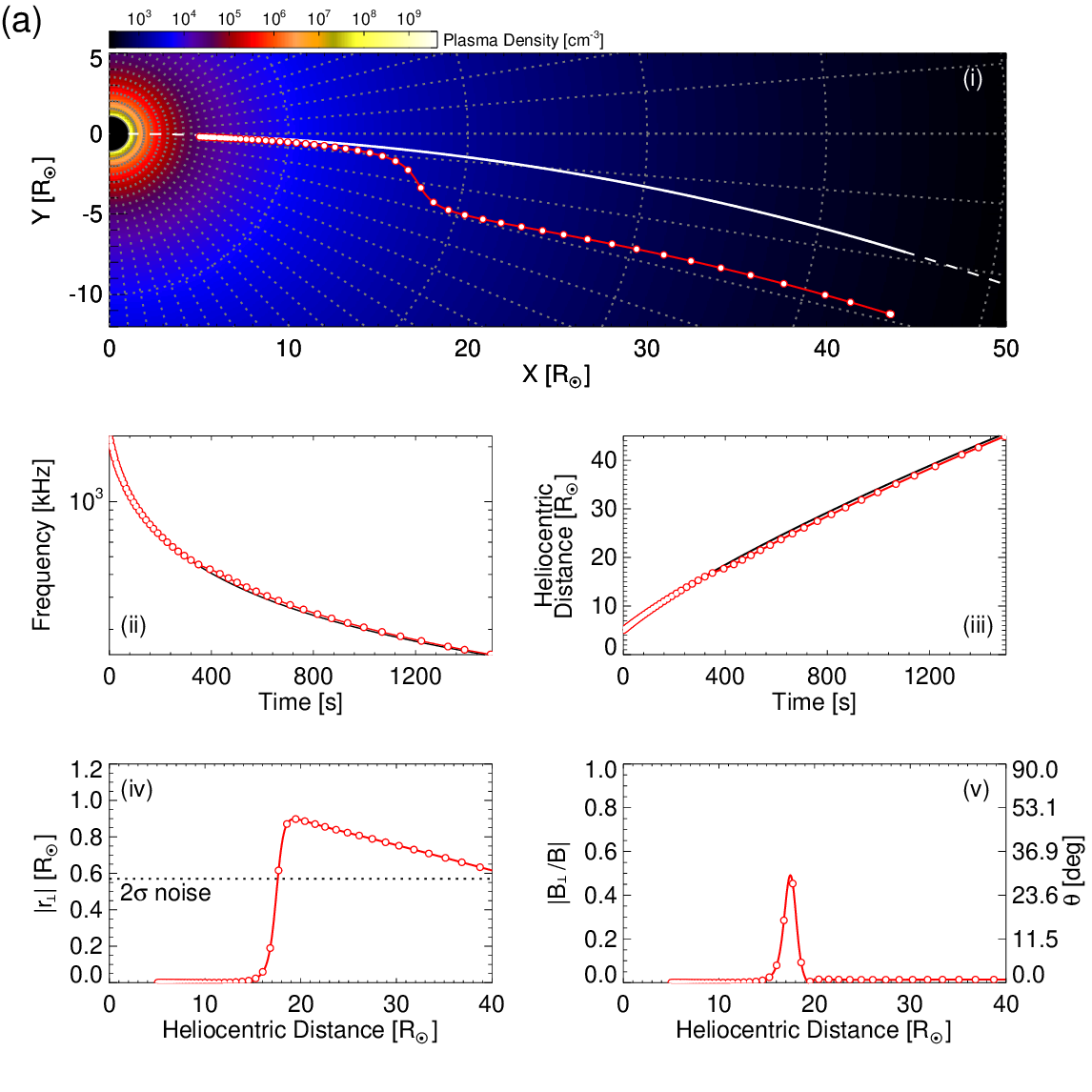}
        \includegraphics[width=0.49\textwidth]{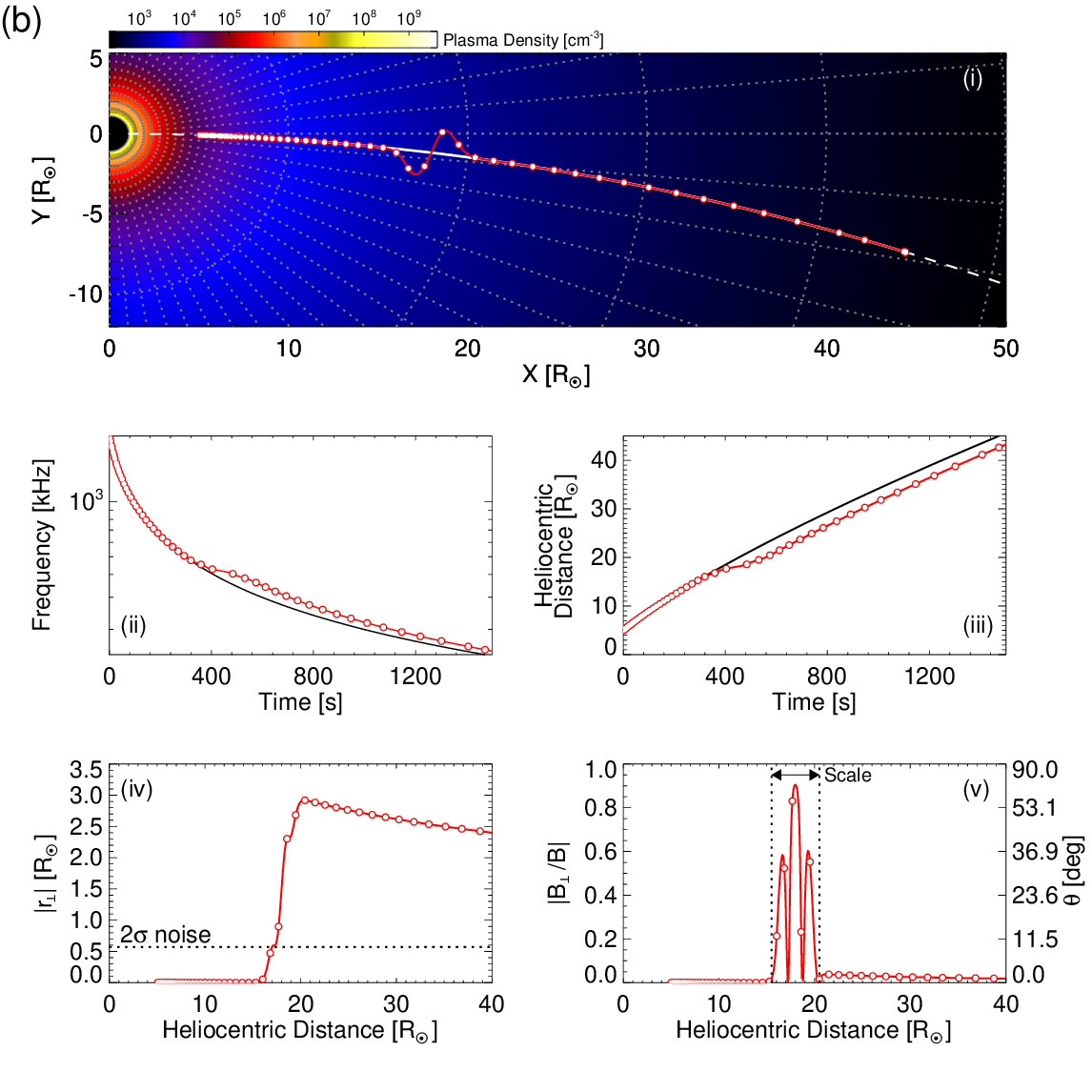}
        \caption{As in figure \ref{fig:kink_Jburst_sim} but for two field perturbations determined by \textit{(a)} an arctan function and \textit{(b)} a double Gaussian function. The upper panels show the field lines for a Parker spiral (white dashed lines) and the perturbed field line (red). The segment of the nominal Parker spiral used to compare to the perturbed field line are shown as solid white lines. The background gradient represents a 2D plasma density field. The lower panels show $f(t)$, $r(t)$, $r_\perp$ and $B_\perp/B$ over time. The dashed black lines represent the $2\sigma$ noise level from Appendix \ref{appendix:noise}. The open red circles in all panels represents the distances probed by PSP due to the spectral resolution of the FIELDS instrument.}
        \label{fig:field_line_sim}
    \end{figure}
    
We further consider two simple disturbances superimposed onto the Parker spiral in Figure \ref{fig:field_line_sim}. The nominal field line is parametrized by $\theta_\mathrm{ps}(r_\mathrm{ps})=-\pi/4\cdot r_\mathrm{ps}/(215\,\mathrm{R}_\odot)$ where $r_\mathrm{ps}$ is the radial distance along the line, shown by the dashed line in the upper panels of Figure \ref{fig:field_line_sim}. 
The red lines include a magnetic field disturbance introduced in angle using (a) an arctan function $\theta_1^\prime$:
    \begin{equation}\label{eq:arctan}
        \theta_1^\prime(r_\mathrm{ps}) = -\frac{A}{r_\mathrm{ps}}\left(1 - \frac{2}{\pi}\arctan\left(-\frac{r_\mathrm{ps}-r_0}{\sigma}\right)\right),
    \end{equation}
    and (b) a double Gaussian $\theta_2^\prime$:
    \begin{equation}\label{eq:dgauss}
        \theta_2^\prime(r_\mathrm{ps}) = -\frac{A}{r_\mathrm{ps}\sigma}\left(\exp\left(-\frac{(r_\mathrm{ps}-r_0)^2}{\sigma}\right) + \exp\left(-\frac{(r_\mathrm{ps}-r_1)^2}{\sigma}\right) \right),
    \end{equation}
where $A$ is set to $2$~R$_\odot$ and $\sigma=1$~R$_\odot$. The disturbances are centred on $r_0=17.5$~R$_\odot$ and $r_1=r_0+1$. 
The final disturbed field lines are then $\theta(r_\mathrm{ps})=\theta_\mathrm{ps}(r_\mathrm{ps})+\theta_{1,2}^\prime(r_\mathrm{ps})$. 
For the disturbance modelled by the arctan function, the method outlined in section \ref{sec:method} estimates a change in $r_\perp$ of almost $1$~R$_\odot$, clearly above the $2\sigma$ noise level, and the gradient of $r_\perp$ leads to an estimate of the field deflection angle which reaches a peak of $\sim30\arcdeg$. Such gradual deflections from the nominal Parker spiral produce only slight changes in the $f(t)$ profile, but demonstrates that even small variations in the burst drift-rate can indicate large-scale field perturbations.
    
The ``kink''-like disturbance resembles a magnetic switchback with a large major axis of several solar radii and a small aspect ratio. Instead of the field folding on itself, the perturbation forms a path that is quasi-perpendicular to the density gradient, causing the drift-rate to decrease in magnitude. Since the direction of each turn along the field line cannot be reconstructed, $r_\perp$ reaches a maximum of $2.9$~R$_\odot$ despite the amplitude of each fluctuation being $2$~R$_\odot$. The gradient of $r_\perp$ produces a complex $B_\perp/B$ profile, reaching up to $65\arcdeg$ over a scale of $\sim5$~R$_\odot$. The red open circles mark the frequencies (distances) corresponding to the spectral resolution of PSP. Whilst PSP data cannot resolve the full $B_\perp/B$ detail, we can estimate two quantities: 
the largest deflection angle from a reference direction (likely underestimated) and the scale of the disturbance. In Figure \ref{fig:field_line_sim}(b), PSP would detect a maximum $\theta$ of $56\arcdeg$, underestimating the true maximum by about $14\%$. This underestimation depends on the height at which the field's maximum deflection occurs for a given event.
    
\section{Observations of Interplanetary Type III Solar Radio Bursts}\label{sec:observations}
        
\begin{figure}[htb!]
        \centering
        \includegraphics[width=0.4\textwidth]{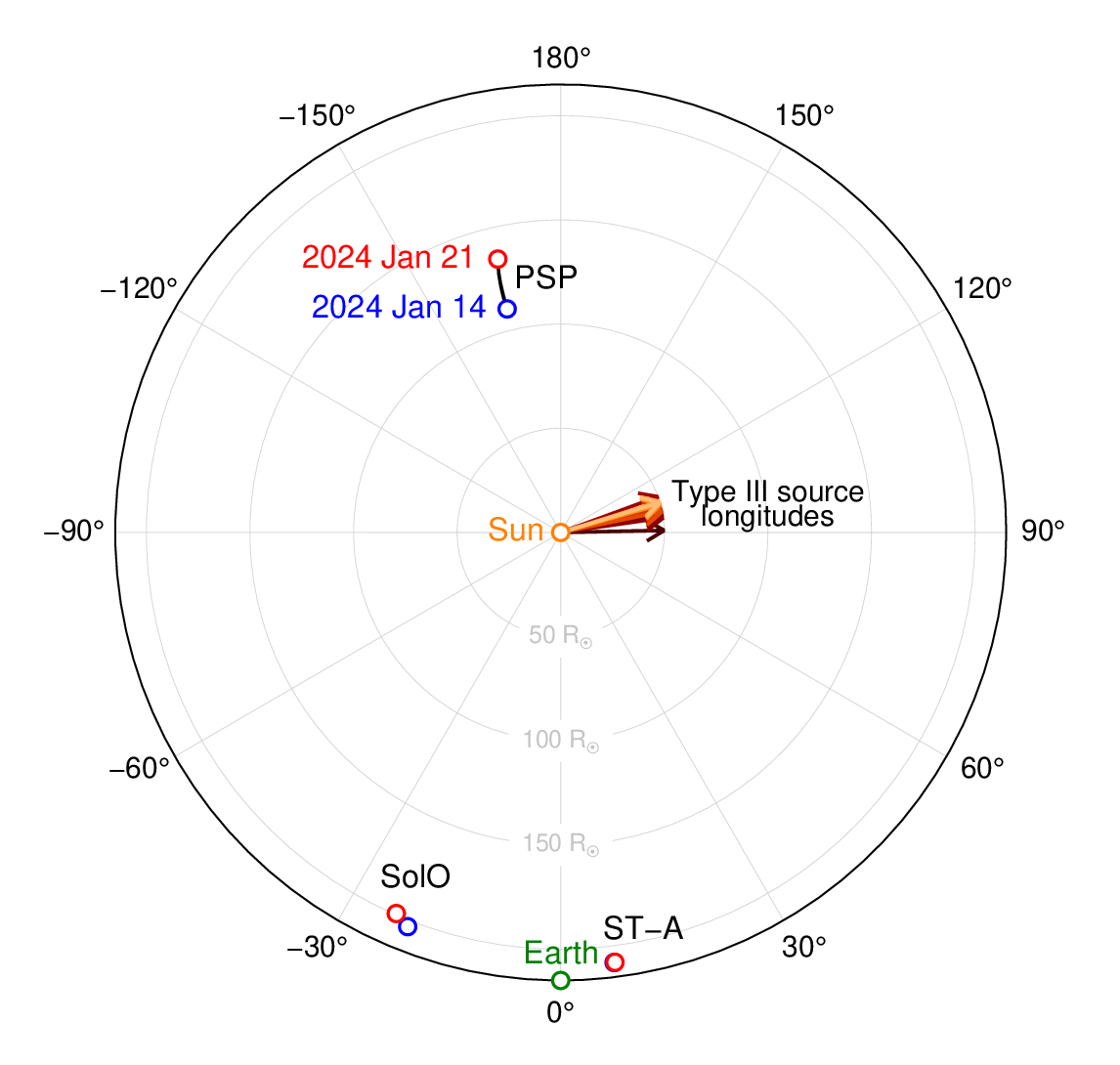}
        \caption{Spacecraft orbits during 14-21 January 2024 (thick black lines) in HEE coordinates. The green circle denotes Earth. Additional spacecraft orbits (Solar Orbiter and STEREO-A) are also shown. The arrows represent the approximate source longitude for 15 of the type III bursts from multi-spacecraft intensity fitting where the colour represents dates increasing from dark to light red.}
        \label{fig:psp_orbit}
    \end{figure}

To assess whether fluctuations in type III radio bursts can be attributed to magnetic field disturbances, we analyse 24 type III radio bursts observed by PSP over a one week period, without selection criteria other than excluding clearly overlapping bursts. We use PSP due to its excellent temporal and spectral resolution.
The events occurred between 14--21 January 2024 and were observed by the Radio Frequency Spectrometer (RFS) onboard PSP/FIELDS \citep{2016SSRv..204...49B}, 
using data from both the High Frequency Receiver (HFR) 
and the Low Frequency Receiver (LFR) with a time cadence of $7$~s. During this orbital period, PSP was at radial distances between 112--133~R$_\odot$ and maintained a heliocentric angle of approximately $-166\arcdeg$ in the HEE coordinate system (Figure \ref{fig:psp_orbit}). 
The burst intensities are reported as a flux density in solar flux units (sfu) where $1\,\mathrm{sfu}=10^{-22}$~W m$^{-2}$~Hz$^{-1}$, and the background level is subtracted. The background level is defined as the median flux density within each frequency channel across a quiet 15 minute interval before or after the burst. 
We restrict the analysis to frequencies between the HFR maximum of $19.2$~MHz, and $100$~kHz, below which the signal-to-noise ratio drops significantly. The approximate source longitudes are localised using an multi-spacecraft intensity fitting approach \citep{2021A&A...656A..34M, 2025NatSR..1511335C} based on the 1~au scaled peak intensities at 1~MHz for 15 of the events that were also observed by the STEREO-A \citep[ST-A;][]{2008SSRv..136..487B} and WIND \citep{1995SSRv...71..231B} spacecraft. The red arrows in Figure \ref{fig:psp_orbit} show that the bursts originate from the same solar region throughout the observing period.

        
    \begin{figure}[htb!]
        \centering
        \includegraphics[width=0.45\textwidth]{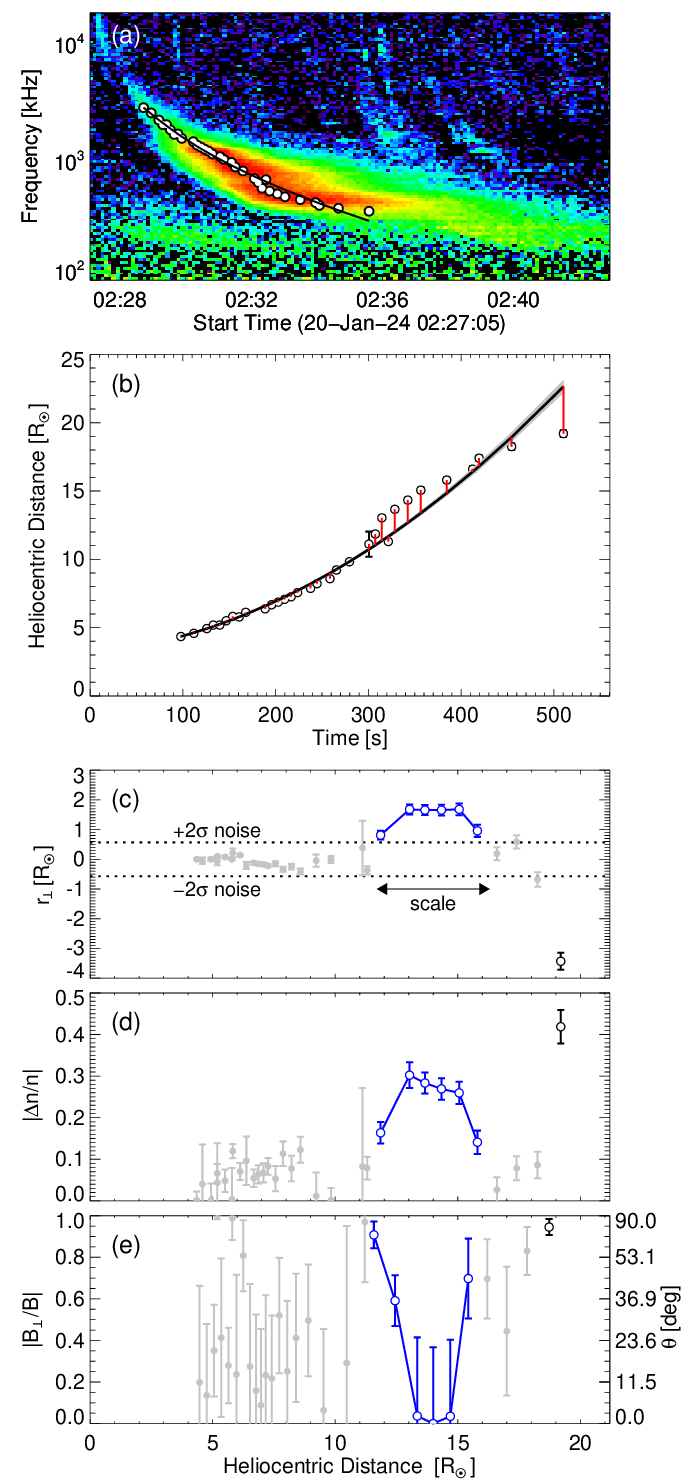}
        \caption{An interplanetary type III burst observed by PSP on 2024 January 20 near 02:29 UT. \textit{(a)} Dynamic spectrum. The white circles mark the time of each intensity peak. \textit{(b)} $r(t)$ profile (white circles). The black line shows the polynomial best fit, and the red lines mark the length of $r_\perp$. The grey region shows the fit corridor from 1000 realisations generated by Monte Carlo sampling of the measurement uncertainties. The black line in panel (a) shows the polynomial best fit converted to frequency space. \textit{(c)} $r_\perp(t)$. The dotted lines represent the $2\sigma$ noise level (Appendix \ref{appendix:noise}). Open circles show values of $r_\perp$ larger than the noise, and closed grey points denote values indistinguishable from noise. Blue circles and connected blue line show a perturbation where a scale can be defined. \textit{(d)} $|\Delta{n}/n|$ over heliocentric distance. \textit{(e)} $|B_\perp/B|$ over heliocentric distance. Since this quantity is determined by the differences in $r_\perp$, the values are plotted against against the midpoint of the two times used for each difference.}
        \label{fig:obs_eg_typeIII}
    \end{figure}
    
To determine $B_\perp/B$ from dynamic spectra, 
we first determine the emitter's distance as a function of time. We trace the burst frequency profile using the timing 
of the peak flux $I_\mathrm{pk}$ of time profile $I(t)$ 
for each frequency channel, as in previous studies
\citep[e.g.][]{1987A&A...173..366D, 2015A&A...580A.137K, Azzollini_2025}. The peak time is used rather than the onset time, which can be ambiguous to define. 
To reduce the influence of poorly resolved intensity peaks,
the peak time is defined as the average time 
across the interval where the flux density is at least 90\% of its maximum. The uncertainty in $I_\mathrm{pk}$ is taken as $\sigma_{I_\mathrm{pk}}=\left[{(0.12I_\mathrm{pk})^2 + I_\mathrm{bg}^2}\right]^{1/2}$, 
combining a $12\%$ signal error ($\sim0.5$~dB 
amplitude resolution \citep{2020A&A...642A..12M}) 
in quadrature with the background level. 
Frequency uncertainties $\sigma_f$ are set 
to half the channel width. 
At higher frequencies, where the drift-rate is large, 
multiple peaks can appear simultaneously 
across adjacent frequency channels. 
For these cases, we compute a weighted average of the frequencies with weights $w_i=1/\sigma^2_{I_\mathrm{pk}}$, 
and define $\sigma_f$ as the standard deviation.
    
To convert the frequencies $f(t)\pm\sigma_f$ to heliocentric distances $r(t)\pm\sigma_r$, we use the density model of equation \ref{eq:Nparker}, assuming fundamental emission. 
To quantify fluctuations, a reference trajectory is required. 
We fit $r(t)$ with a quadratic polynomial
    \begin{equation}\label{eq:polynomial}
        r_\mathrm{fit}(t) = r_0 + vt + \frac{1}{2}at^2,
    \end{equation}
which assumes constant acceleration following $v(t)=v_0 + at$, where $v_0$ is the initial velocity at $t=0$ and $r_0$ is the initial heliocentric distance 
\citep[e.g.][]{2015A&A...580A.137K}. 
The residuals then provide an estimate of the perpendicular distance from the radial reference path: $r_\perp \simeq r(t)-r_\mathrm{fit}(t)$. The change in $r_\perp$ between successive points gives $\Delta r_\perp$, while $\Delta r$ is the corresponding change in the radial distance. From these these two quantities, $\tan\theta$ is approximated as $\tan\theta\approx\Delta r_\perp/\Delta r$, from which $B_\perp/B$ can be evaluated.

We also assess the fluctuations in $r(t)$ assuming they are caused by density variations. The fitted model from equation \ref{eq:polynomial} is converted to frequency-space using equation \ref{eq:Nparker}, and the frequency deviation $\Delta{f}$ is estimated as $\Delta{f}\simeq f(t)-f_\mathrm{fit}(t)$. The corresponding density fluctuation ratio is approximated as $\Delta{n}/n\simeq2\Delta{f}/f_\mathrm{fit}$, based on the relationship between plasma frequency and the square-root of the electron density.

Figure \ref{fig:obs_eg_typeIII} shows an example of an interplanetary type III radio burst observed by PSP on 2024 January 20, starting near 02:28 UT. The identified intensity peaks span approximately 5~minutes, drifting from $2.97$~MHz (4.35~R$_\odot$) to 388~kHz (19.2~R$_\odot$). A change in the observed drift-rate is considered to be associated with a real disturbance if at least one value of $r_\perp$ exceeds the $2\sigma$ noise level (Appendix \ref{appendix:noise}). For this event, $r_\perp$ deviates from the best-fit curve between (300--400)~s, reaching $1.7$~R$_\odot$ over a scale of (4--5)~R$_\odot$. To quantify the scale of each disturbance, we measure the distance between the first point where $r_\perp$ exceeds the noise level and the nearest point where it returns below. In Figure \ref{fig:obs_eg_typeIII}, a singular data point has a large value of $|r_\perp|$ after 500~s, for which a scale cannot be determined. We also include a fit corridor generated by Monte Carlo sampling of the data uncertainties, performing 1000 realisations in which the measurements were perturbed within their errors using values drawn from a normal distribution and the polynomial $r_\mathrm{fit}(t)$ refitted each time. 
The bulk of the collated residuals lie within the propagated uncertainty, indicating that the error estimates 
are a reliable measure.
    
From the polynomial fit, the burst propagates with speeds between $0.07c$ to $0.02c$. These are lower bounds, assuming the beam follows a perturbed, non-radial path. The fit also implies constant acceleration rather than deceleration due to energy losses, which likely reflects the effect of perturbations on the best-fit rather than a physical energy gain. Panels (d) and (e) show that the perpendicular motion can be explained either by a $\pm30\%$ change in density or a magnetic field disturbance where the field trajectory is perturbed by $\theta=(65.5\pm11)\arcdeg$ from the radial direction ($|B_\perp/B| = 0.91\pm0.06$). 
In Figure \ref{fig:obs_eg_typeIII}, $B_\perp/B$ falls to almost zero soon after the initial perturbation. This occurs when $\Delta r_\perp$ changes little between successive points, suggesting that the field is first displaced from a reference position, and then follows a quasi-parallel path over some distance, before bending back toward the reference trajectory.
    
\begin{figure}[htb!]
    \centering
    \includegraphics[width=1\textwidth]{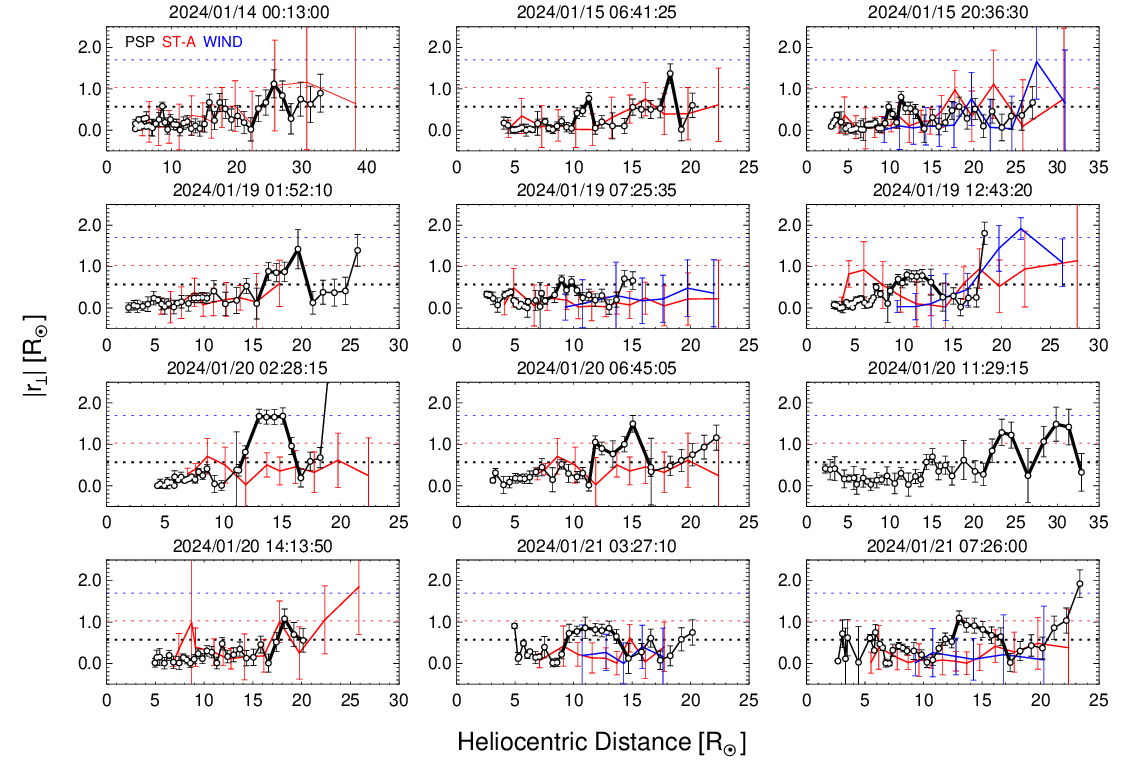}
    \caption{$|r_\perp|$ determined as a function of distance used as a proxy for the magnetic field line structure (the direction of each perpendicular displacement is not determined). The 12 events observed by PSP that presented disturbances above the noise level (dotted lines) are shown, with the corresponding disturbances highlighted by thicker line segments. Fluctuations outside these segments is considered as noise. Black lines are data from PSP. Red and blue lines represent $|r_\perp|$ for the same bursts using ST-A and WIND data where the burst is visible.}
    \label{fig:field_paths_proxy}
\end{figure}
    
Across all events observed by PSP, 50\% show fluctuations that rise above the $2\sigma$ noise level, with an average and median $|r_\perp|$ of $\sim1.1$~R$_\odot$ (Figure \ref{fig:scales}). Figure \ref{fig:field_paths_proxy} shows $|r_\perp|$ as a function of distance for these 12 events, which can be considered a proxy for the magnetic field structure, yet we recall that the direction of each perpendicular displacement is not determined. The white circles connected by thin lines indicate fluctuations consistent with noise, while the thick line segments mark deviations from the reference trajectory. The scales of these disturbances are summarised in Figure \ref{fig:scales}. For the selected type III bursts, the method probes scales between (1.8--6.4)~R$_\odot$, at distances from (9--30)~R$_\odot$. Panels (b) and (c) show the maximum value of $\Delta{n}/n$ and $B_\perp/B$ within each disturbance. Across the events, $\Delta{n}/n$ averages $17\%$ (median 17\%) and ranges between $10\%$ and $30\%$. The average change in field angle is $47\arcdeg$ (median $54\arcdeg$), with a range of (23--88)$\arcdeg$. For the PSP observed bursts that presented fluctuations above the $2\sigma$ level (shown in Figure \ref{fig:field_paths_proxy}), we also perform the same analysis using data from ST-A/WAVES HFR and WIND/WAVES RAD1 receivers. For these spacecraft, the larger temporal and spectral resolutions lead to larger noise estimates by a factor of 2 and 3 times that of PSP (Appendix \ref{appendix:noise}), and are unable to identify fluctuations above a $2\sigma$ level.

\section{Signatures of Field Disturbances in Type III Radio Bursts}\label{sec:typeIII_sim}

We next investigate whether a switchback-like field perturbation can produce additional observational features in type III burst spectra. This is done using a numerical simulation of an electron beam distribution evolving in time and space and the subsequent generation of Langmuir waves and radio emission.
    
\subsection{Numerical Simulation Description}\label{sec:num_desc}

The evolution of an electron beam distribution $f(v,r,t)$~[cm$^{-4}$ s] and the spectral energy density of 
Langmuir waves $W(v,r,t)$~[erg cm$^{-2}$] 
is computed using the code by \citet{2001SoPh..202..131K, 2001CoPhC.138..222K} solving one-dimensional kinetic equations:
    \begin{equation}\label{eq:Frvt}
            \frac{\partial{f}}{\partial{t}} + v\frac{\partial{f}}{\partial{r}} = \frac{4\pi^2 e^2}{m_\mathrm{e}^2} \frac{\partial}{\partial{v}}\left(\frac{W}{v}\frac{\partial{f}}{\partial{v}}\right)\;,
    \end{equation}
    \begin{equation}\label{eq:Wrvt}
            \frac{\partial{W}}{\partial{t}} + \frac{\partial{\omega}}{\partial{k_\mathrm{L}}}\frac{\partial{W}}{\partial{r}} - \frac{\partial{\omega_\mathrm{pe}}}{\partial{r}}\frac{\partial{W}}{\partial{k_\mathrm{L}}} = \frac{\pi\omega_\mathrm{pe}}{n_\mathrm{e}}v^2 W \frac{\partial{f}}{\partial{v}} - \gamma W,\;\;\; \omega_\mathrm{pe}=k_\mathrm{L}v,
    \end{equation}
where $m_e$ and $e$ are the electron mass and charge, $\omega_\mathrm{pe}$ is the plasma frequency, and $k_\mathrm{L}$ is the wavenumber of the Langmuir waves. The term $\gamma W$ represents the Landau damping of the plasma waves, with $\gamma=\pi n e^4 z_p^2 \ln\Lambda /m_\mathrm{e}^2 v_\mathrm{Te}^3$ where $\ln\Lambda\approx20$ is the Coulomb logarithm, $z_p=1.18$ gives the average atomic number in the photosphere \citep{2011A&A...536A..93J}, $v_\mathrm{Te}=\sqrt{k_\mathrm{B}T/m_\mathrm{e}}$ is the thermal velocity for temperature $T$, and $k_\mathrm{B}$ is the Boltzmann constant. Equations \ref{eq:Frvt} and \ref{eq:Wrvt} interact resonantly under the condition $\omega_\mathrm{pe}=k_\mathrm{L}v$, on the quasi-linear timescale $\tau_\mathrm{ql}(r)=n(r)/(n_\mathrm{b}\omega_\mathrm{pe}(r))$. 

\begin{figure}[b!]
    \centering
    \includegraphics[width=0.45\textwidth]{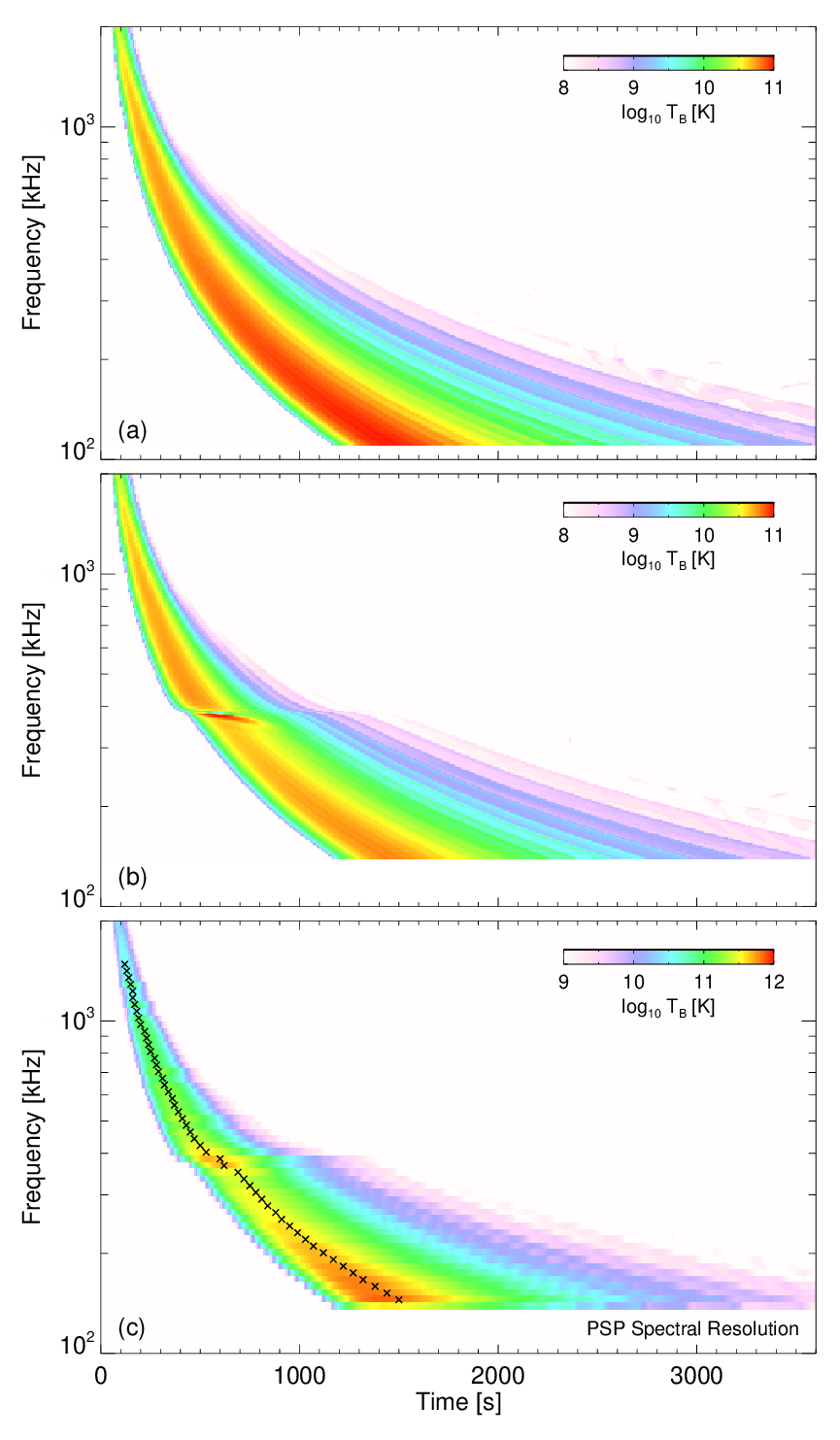}
    \caption{Dynamic spectra from  the numerical simulations of type III radio burst. \textit{(a)} Using a radial magnetic field without a field disturbance. \textit{(b)} Using a perturbed magnetic field with a kink fluctuation as shown in Figure \ref{fig:field_line_sim}(b). \textit{(c)} As in panel (b) but reduced in spectral resolution to emulate that of PSP. The black crosses trace the peak intensity along the spine of the burst.}
    \label{fig:typeIII_sim}
\end{figure}
        
The initial distribution of electrons is given by
\begin{equation}
        f(v,r,t=0) = \frac{v}{v_0}\frac{2n_\mathrm{b}}{v_0}\exp{\left(-\frac{r^2}{d^2}\right)},\,\, 0<v<v_0
\end{equation}
where $n_\mathrm{b}=2\times10^{-3}$~cm$^{-3}$ is the beam density, $d=2\times10^{10}$~cm is the spatial size, and $v_0=3\times10^9$~cm s$^{-1}\simeq0.1c$ is the initial velocity. At $t=0$, the distribution is located at $r_0=1.1\times$10$^{11}$~cm ($2.6$~R$_\odot$). The background spectral energy density is set to $W(r,v,t=0) = 10^{-8} \approx m_\mathrm{e}n_\mathrm{b}v_\mathrm{b}^3/\omega_\mathrm{pe}$. Simulated dynamic spectra are constructed by considering the saturation level of the plasma emission process $L\rightarrow T+S$ \cite[e.g.][]{1995lnlp.book.....T, 2017SoPh..292..117L} where $L$, $T$ and $S$ refer to Langmuir waves, transverse electromagnetic waves, and ion sound waves with angular frequencies $\omega_\mathrm{L}$, $\omega_\mathrm{T}$ and $\omega_\mathrm{S}$. The process saturates when $\omega_\mathrm{s} \gg \omega_\mathrm{L}$, and assuming that $\omega_\mathrm{T}\approx\omega_\mathrm{L}$ and $k_\mathrm{S}\approx k_\mathrm{L}$, and the brightness temperature $T_\mathrm{B}$ is approximated as $T_\mathrm{B}\approx (2\pi)^2v/\omega_\mathrm{pe}\cdot W(v,r,t)$ \citep{2018ApJ...867..158R, 2021NatAs...5..796R}. Taking the peak $T_\mathrm{B}$ at each distance for a given time provides a brightness temperature at each frequency, which is then used to construct the dynamic spectra.
        
\subsection{Emulating a Field Deviation using Kinetic Simulations}

We perform numerical simulation both with and without a magnetic field disturbance. Since the simulations are one-dimensional, a two-dimensional disturbed magnetic field line is emulated by modifying the density gradient with
        \begin{equation}\label{eq:nmod}
            n_\mathrm{mod}(r)=n(r) \times \left(1 + \xi(r) \right)\,,
        \end{equation}
        where
        \begin{equation}
            \xi(r)=\eta \left( 1 - \frac{2}{\pi}\arctan{\left( -\frac{r - r_\mathrm{c}}{h} \right)} \right)\,,
        \end{equation}
where, $\eta=0.3$ and $h=1.5$~R$_\odot$ are the parameters controlling the extent of the density enhancement around $r_\mathrm{c}$. 
The modified density structure approximates the density experienced by an emitter propagating 
along the perturbed field line shown in Figure \ref{fig:field_line_sim}b.

\subsection{Observational Spectro-Temporal Features in Dynamic Spectra}\label{sec:obs_features}

Figure \ref{fig:typeIII_sim} shows the dynamic spectra from the numerical simulations. Panel (a) presents the type III burst produced by electrons propagating along a radial magnetic field without a disturbance, producing a smoothly decreasing drift-rate with no spectral features. Panel (b) presents the type III burst generated by electrons propagating along a field line with a kink disturbance. Compared with the analytical $f(r)$ profile (Section \ref{sec:Bfluc_sim}), the numerical simulations reproduce the reduction in drift-rate, 
which also manifests as a visible time delay along the onset and decay separating the burst into two branches. 
Two additional observational signatures emerge: a break in the intensity along the burst spine and an intensity enhancement that produces stria-like fine structures, similar to that observed
\citep[e.g.][]{2020ApJS..246...49P,2025ApJ...985L..27K}.
The break arises from the change in the density gradient 
along the path, which suppresses the Langmuir-wave generation, and is a consequence of the density inhomogeneity term in equation \ref{eq:Wrvt} \citep[see, for example,][]{2001A&A...375..629K}. The intensity enhancement occurs because emission is maintained near a fixed plasma frequency (distance) for an extended period due to the structure of the field deviation.
        \begin{figure}[htb!]
            \centering
            \includegraphics[width=0.8\textwidth]{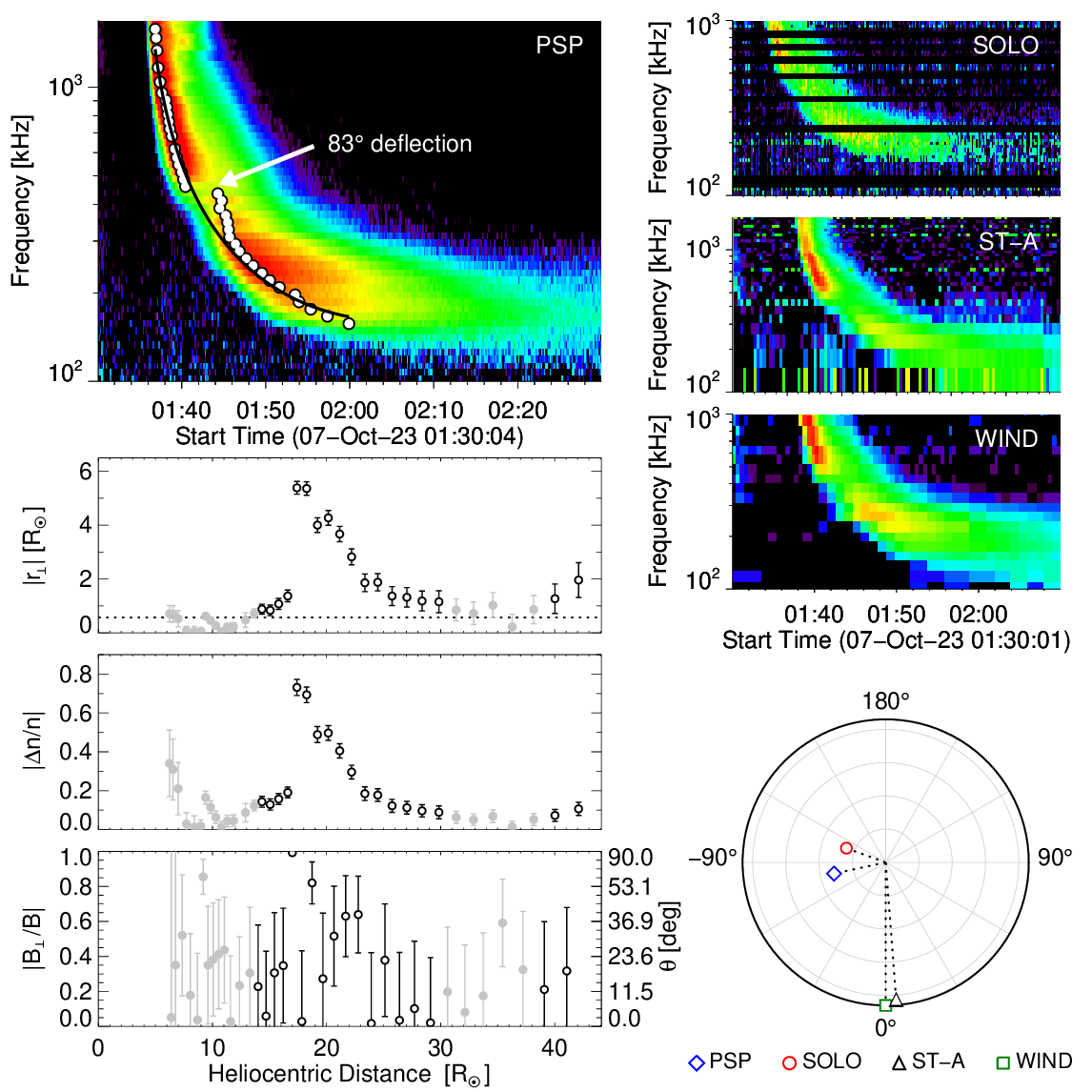}
            \caption{A type III burst observed by PSP near 01:40 UT on 2024-Oct-7 that presents features similar to those in Figure \ref{fig:typeIII_sim}. The lower panels present the same analysis applied to the burst in Figures \ref{fig:obs_eg_typeIII} and \ref{fig:field_paths_proxy}. Dynamic spectra observed by SolO, ST-A, and WIND are shown (prominent bands of RFI are removed from SolO spectra), along with the spacecraft positions in the HEE coordinate system where the outer boundary represents 1~au.}
            \label{fig:typeIII_features1}
        \end{figure}
        Panel (c) shows the same spectrum adjusted to the spectral resolution of PSP. The localised drift-rate reduction and time delays remain, whilst the intensity break disappears and the intensity enhancement broadens, highlighting that some features may not be present dependent on the scale of the disturbance with respect to the instrumental capability.
        
        Using these observational characteristics as a guide, Figure \ref{fig:typeIII_features1} presents an observational example of a type III burst with similar features that could have been produced whilst the emitter propagated along a magnetic field disturbance. We also include three additional examples in Appendix section \ref{appendix:additional_burst_features}. The bursts are analysed in the same way as outlined in section \ref{sec:observations}. However, since the bursts are identified to have clear segments with reduced drift-rates, leading to $r(t)$ profiles that do not follow a simple quadratic, the polynomial fits to $r(t)$ exclude the few data points attributed to the perturbation. This ensures a clear reference curve with which to measure $r_\perp$ from. For each burst, the measured value of $|r_\perp|$ is several solar radii, yet we remind that since we do not know the direction of each perturbation, the field itself could oscillate similar to that shown in Figure \ref{fig:field_line_sim}b. Nevertheless, these clear perturbations occur across scales near (5--15)~R$_\odot$. In each example, the drift-rate reduction is stronger than in the 24 bursts analysed in section \ref{sec:observations}, and would require unrealistically large density fluctuations (between 50\% and 100\%) \textit{along} the magnetic field to explain. This suggests that the most plausible cause of the observed features is a deflection of the magnetic field line. In each example, we also show the dynamic spectra from other spacecraft. In most cases, the intensity break appears in all observations. Fine structures, however, are harder to detect and depend on the spectral and temporal resolution, as well as the level of RFI. In Figure \ref{fig:typeIII_features4}, PSP and WIND show complex, overlapping type III bursts, whilst in the analysed ST-A spectra, these additional bursts are either not visible or much fainter than the main burst. As a result, it is difficult to determine the extent to which a given feature is altered by overlapping emission.
        
\section{Summary}\label{sec:summary}
    
The results of this study show fluctuations 
in the drift-rate of interplanetary type III radio bursts 
and provide interpretation in terms of disturbed magnetic fields. 
In 50\% of selected events observed by Parker Solar Probe throughout one week, 
the burst frequency drift-rates vary in ways that cannot be convincingly 
explained by motion along a radial trajectory through a spherically symmetric 
and monotonically decreasing plasma density. 
We investigated whether disturbances in the guiding magnetic field, rather than density fluctuations along the field line, could account for these variations 
and identified the signatures such disturbances leave in type III dynamic spectra. Perturbations were detected over scales of (1.8--6.4)~R$_\odot$ at heliocentric distances between (9--30)~R$_\odot$. 
Explaining the perturbations purely via density 
fluctuations parallel to the field path would 
require fluctuations between (10--30)\%, 
whilst to explain as local magnetic field deviations requires angular changes along the field path between (23--88)$\arcdeg$.
    
Simulations of magnetic field disturbances resembling large-scale switchbacks with small aspect ratios---where the perturbed path is quasi-perpendicular to the radial density gradient---shows that the major axis of such disturbances is detectable within the spectral resolution of Parker Solar Probe. Switchbacks with high aspect ratios, where the field direction is reversed, could cause the type III drift-rate to flip sign, though this will be challenging to observe with current spectral limitations.
    
Our method to infer magnetic field disturbances from type III bursts rests on several assumptions. We assumed fundamental emission; if the bursts were harmonic, both the height and scale of each disturbance would differ, 
and the inferred profile of $B_\perp/B$ would change. 
To quantify the fluctuations, a reference direction is required. We used a quadratic fit to the observations because it produced small residuals, though earlier studies have used power-law fits \citep{2015A&A...580A.137K, Azzollini_2025}. 
For the range of frequencies and durations investigated 
in this work, we found each model produced virtually indistinguishable results; hence, the choice of fit does not affect our conclusions. We further note that intensity variability, modelled in section \ref{appendix:noise}, adds uncertainty to the timing of the radio peak. This is partly dependent on the temporal resolution of the observing instrument, and future improvements are essential to providing robust constraints on the peak timing and the resulting drift-rate variability.
    
Numerical simulations further indicate that magnetic switchback-like disturbances produce distinct observable spectral 
features: a localised reduction of the frequency drift-rate, which also appears as delayed emission separating the burst into two branches; suppressed radio emission due to changes in the density gradient along the perturbed path; 
and enhanced emission resembling stria fine structures caused by sustained emission within the same frequency channel. The appearance of fine structures for radio emission along disturbed field structures offers a novel explanation for the generation of striae in interplanetary bursts, in addition to the extension of the coronal density turbulence mechanism. Guided by these characteristics, we identify four examples of type III bursts exhibiting some or all of these features. Analysis suggests that these features are unlikely to result from density changes, which would need to be unrealistically large along the field path, and are more plausibly explained by deflections of the magnetic field. Another potential observational signature is a reduction in the magnitude of the circular polarisation during the disturbance, as changes in the electron beam direction can affect the sense of o-mode radiation. 

The results suggest that magnetic field fluctuations can leave a detectable signature in the dynamic spectra of type III radio bursts and should be considered alongside density fluctuations. 
In several cases, large magnetic field deviations are required in order to explain the features in the dynamic spectra. These findings further emphasize that radio observations can serve as a valuable and unique diagnostic of inner heliospheric conditions, offering spatial coverage where in-situ measurements remain limited.

\begin{acknowledgments}
This work was supported by STFC/UKRI grant ST/Y001834/1.
E.P.K. is supported by the Leverhulme Trust (Research Fellowship RF-2025-357). 
The authors thank the PSP/RFS team for making the data available. The FIELDS experiment on the Parker Solar Probe spacecraft was designed and developed under NASA contract NNN06AA01C. This research has made use of the Astrophysics Data System, funded by NASA under Cooperative Agreement 80NSSC21M00561. 
\end{acknowledgments}

\clearpage
\appendix
\counterwithin{figure}{section}
\renewcommand\thefigure{\thesection\arabic{figure}}

\section{Characteristic scales of disturbances}
\begin{figure}[htb!]
        \centering
        \includegraphics[width=0.45\textwidth]{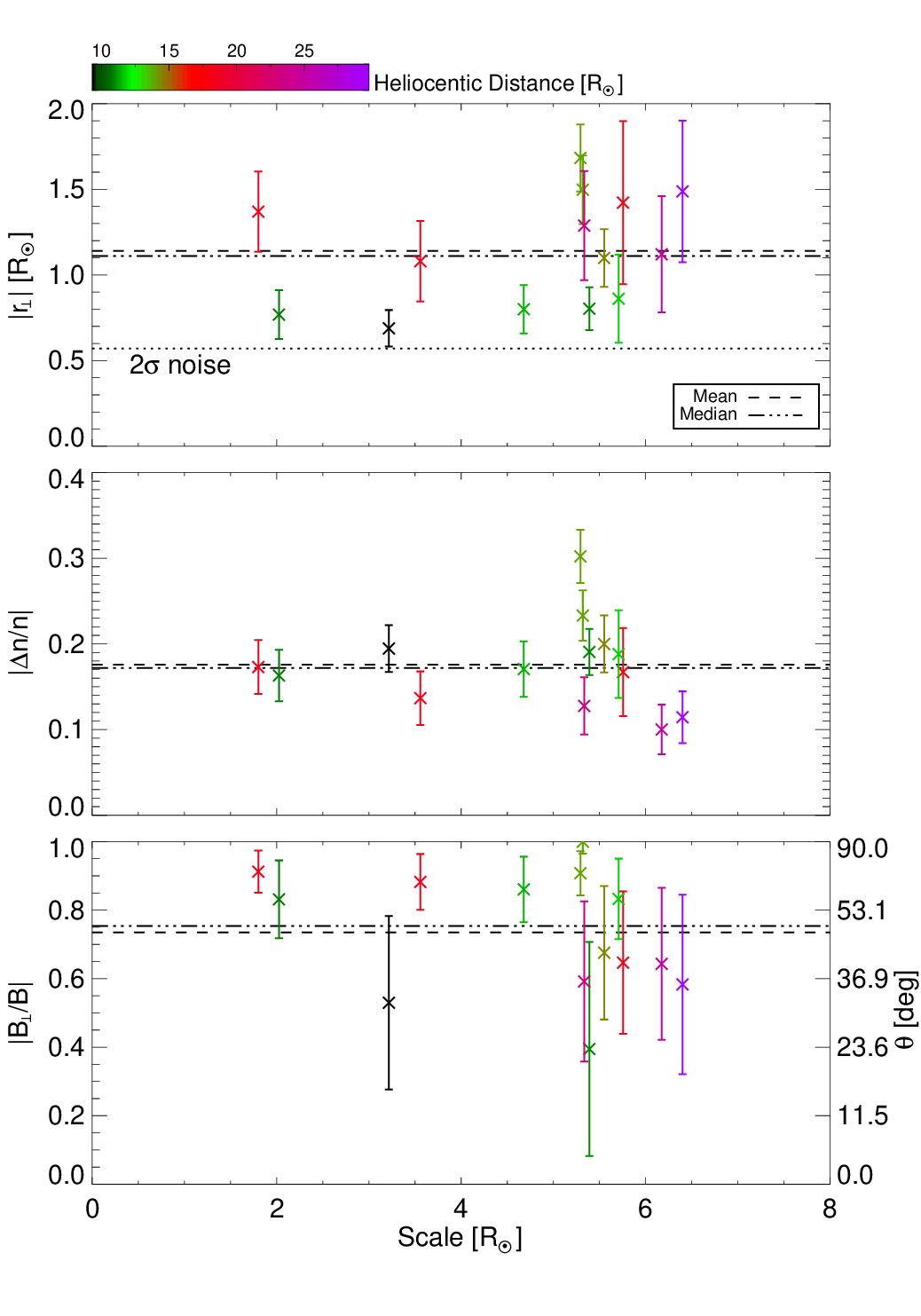}
        \caption{Scales and distances of individual disturbances in density and magnetic field throughout the 24 type III burst events. 
        The dashed lines show the mean and median value across all scales.}
        \label{fig:scales}
    \end{figure}

The appendix summarises the analysis of the type III solar radio bursts.
Figure \ref{fig:scales} presents the maximum values of $|r_\perp|$, $|\Delta{n}/n|$ and $|B_\perp/B|$ within each disturbance against the scale of the disturbance.

\section{Uncertainty estimate}\label{appendix:noise}

\begin{figure}[htb!]
    \centering
    \includegraphics[width=0.4\textwidth]{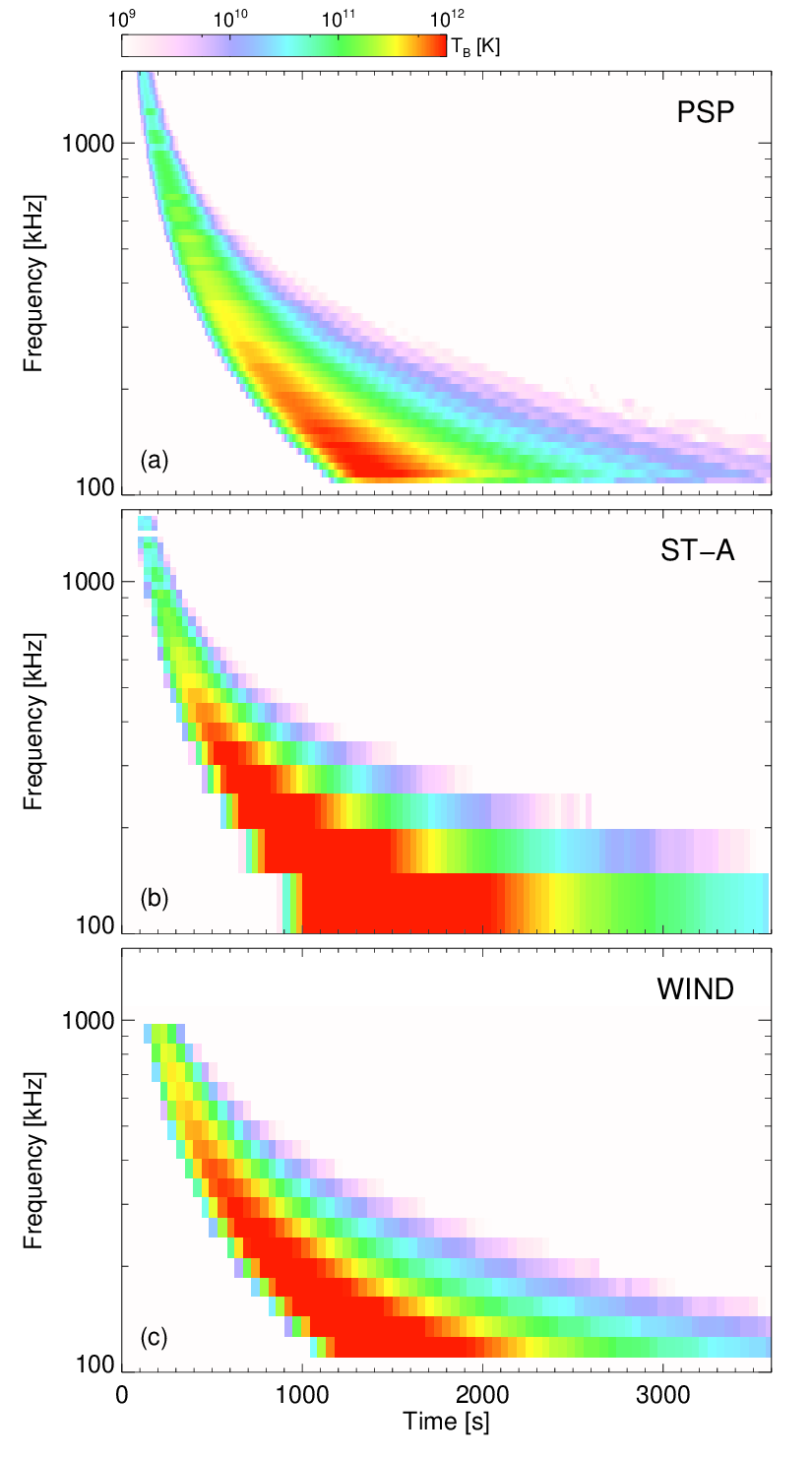}
    \caption{Simulated dynamic spectra reduced to the temporal and spectral resolutions of (a) PSP (b) ST-A, and (c) WIND.}
    \label{fig:typeIII_sim_psp_sta_wind}
\end{figure}

\begin{figure}[htb!]
    \centering
    \includegraphics[width=0.5\textwidth]{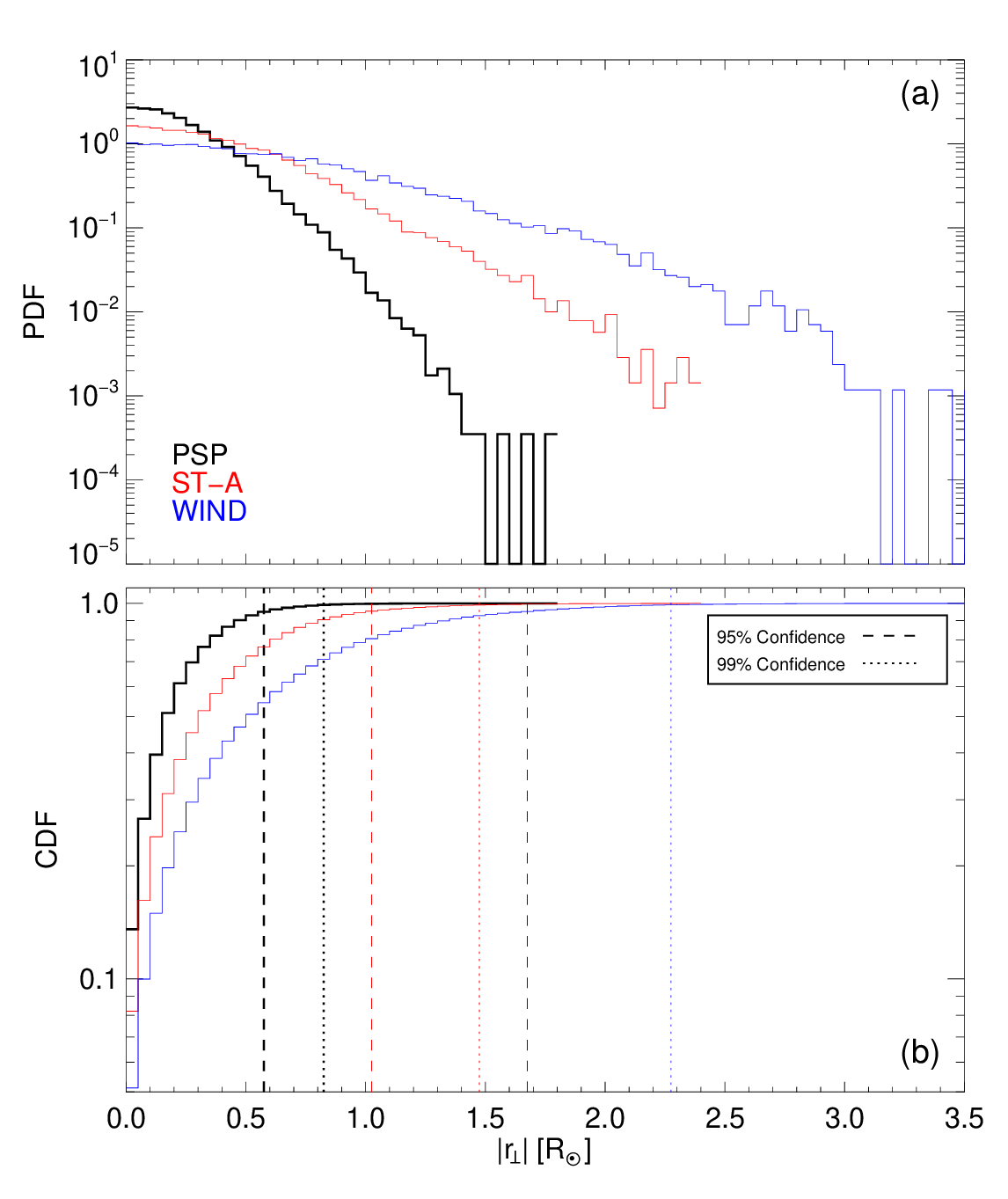}
    \caption{\textit{(a)} Probability density of the collated values of $r_\perp$ from each realisation of the perturbed simulation data for dynamic spectra constructed using PSP (black), ST-A (red), and WIND (blue) temporal and spectral resolutions.} \textit{(b)} Cumulative distribution of the collated values of $r_\perp$. The dashed and dotted lines mark the scales of $r_\perp$ below which 95\% ($2\sigma$) and 99\% ($3\sigma$) of the data exist.
    \label{fig:noise}
\end{figure}

To estimate the uncertainties due to the temporal resolution and uncertainty in the peak intensity, we use a simulated type III burst with a radial magnetic field (Figure \ref{fig:typeIII_sim}a) reduced to the temporal and spectral resolution of PSP, ST-A, and WIND (Figure \ref{fig:typeIII_sim_psp_sta_wind}).
We apply a stochastic Monte Carlo approach, generating 1000 realisations of the dynamic spectrum by perturbing each data point by an amount within $\pm\Delta t/2$ and $\pm25\%$ intensity, with the pseudo-random perturbations drawn from a uniform distribution. The intensity variation reflects typical observational differences between spacecraft on the same radial line \citep[see Figure 5 of][]{2025NatSR..1511335C}. For each realisation, we find $r_\perp$ as outlined in section \ref{sec:method}. The probability density of the collection of $r_\perp$ values for all realisations is shown in Figure \ref{fig:noise}(a), and the cumulative density in panel (b). We take the noise level at the $2\sigma$ level where 95\% of the values exist below $0.57$~R$_\odot$ for PSP, 1.03~R$_\odot$ for ST-A, and $1.68$~R$_\odot$ for WIND.

\section{Uncertainty Propagation}\label{appendix:uncert}

    The uncertainty on the measured values of $r_\perp$ is given by $\sigma_{r_\perp}(t_i)=\left[{\sigma_r(t_i)^2 + \sigma^2_{r_\mathrm{fit}}(t_i)}\right]^{1/2}$ for a given observed time $t_i$. Here, $\sigma_{r_\mathrm{fit}}$ is the uncertainty in the evaluated fit $r_\mathrm{fit}$ and given by $\sigma_{r_\mathrm{fit}}=\sqrt{\mathbf{J}^\mathrm{T} \Sigma \mathbf{J}}$, where $\Sigma$ is the covariance matrix returned by the fit routine, $\mathbf{J}=\left[1, t, (1/2)t^2\right]$ is the Jacobian vector for the quadratic polynomial, and $\mathbf{J}^\mathrm{T}$ is its transpose. The uncertainty on $\Delta{r_\perp}$ and $\Delta{r}$ ($\sigma_{\Delta r_\perp}$ and $\sigma_{\Delta r}$, respectively) are calculated by summing in quadrature the uncertainties on successive values of $r_\perp$ and $r$. The uncertainty on $B_\perp/B$ is given as $\sigma_{B_\perp/B} = \left|\partial{(B_\perp/B)}/\partial{(\tan\theta)}\right|\sigma_{\tan\theta}$. Since $\tan\theta\approx\Delta r_\perp/\Delta r$, then $\sigma_{\tan\theta}=1/|\Delta r| \sqrt{\sigma_{\Delta r_\perp}^2 + \tan^2\theta \cdot \sigma_{\Delta{r}}^2}$, and $\partial{(B_\perp/B)}/\partial{(\tan\theta)}=1/(1+\tan^2\theta)^{3/2}$. The uncertainty in $B_\perp/B$ is smaller when the perpendicular displacement $\Delta{r_\perp}$ is large compared to its measurement error. Points with small $\Delta{r_\perp}$ relative to their uncertainties have correspondingly larger errors. Physically, this indicates that a strong, well-defined field deviation producing a large perpendicular displacement can be determined more reliably than a broad or weak deviation involving only a small deflection.

    Uncertainty in the density fluctuation ratio $\Delta{n}/n=2\Delta{f}/f_{\mathrm{fit}}=2(f-f_{\mathrm{fit}})/f_{\mathrm{fit}}$ is propagated via the conversion of the polynomial fit $r_\mathrm{fit}$ to frequency space to gain $f_\mathrm{fit}$, together with the uncertainty in the frequency $\sigma_f$, taken as $\sigma_{\Delta{n}/n} = \left[\left(\frac{\partial(\Delta{n}/n)}{\partial f}\sigma_f\right)^2 + \left(\frac{\partial(\Delta{n}/n)}{\partial f_{\mathrm{fit}}}\sigma_{f_{\mathrm{fit}}}\right)^2\right]^{1/2}$, where $\sigma_{f_{\mathrm{fit}}}=|\mathrm{d}f/\mathrm{d}r|\sigma_{r_{\mathrm{fit}}}$.

\section{Additional Examples of Type III Bursts with Field Disturbance Features}\label{appendix:additional_burst_features}

\begin{figure}[htb!]
        \centering
        \includegraphics[width=0.8\textwidth]{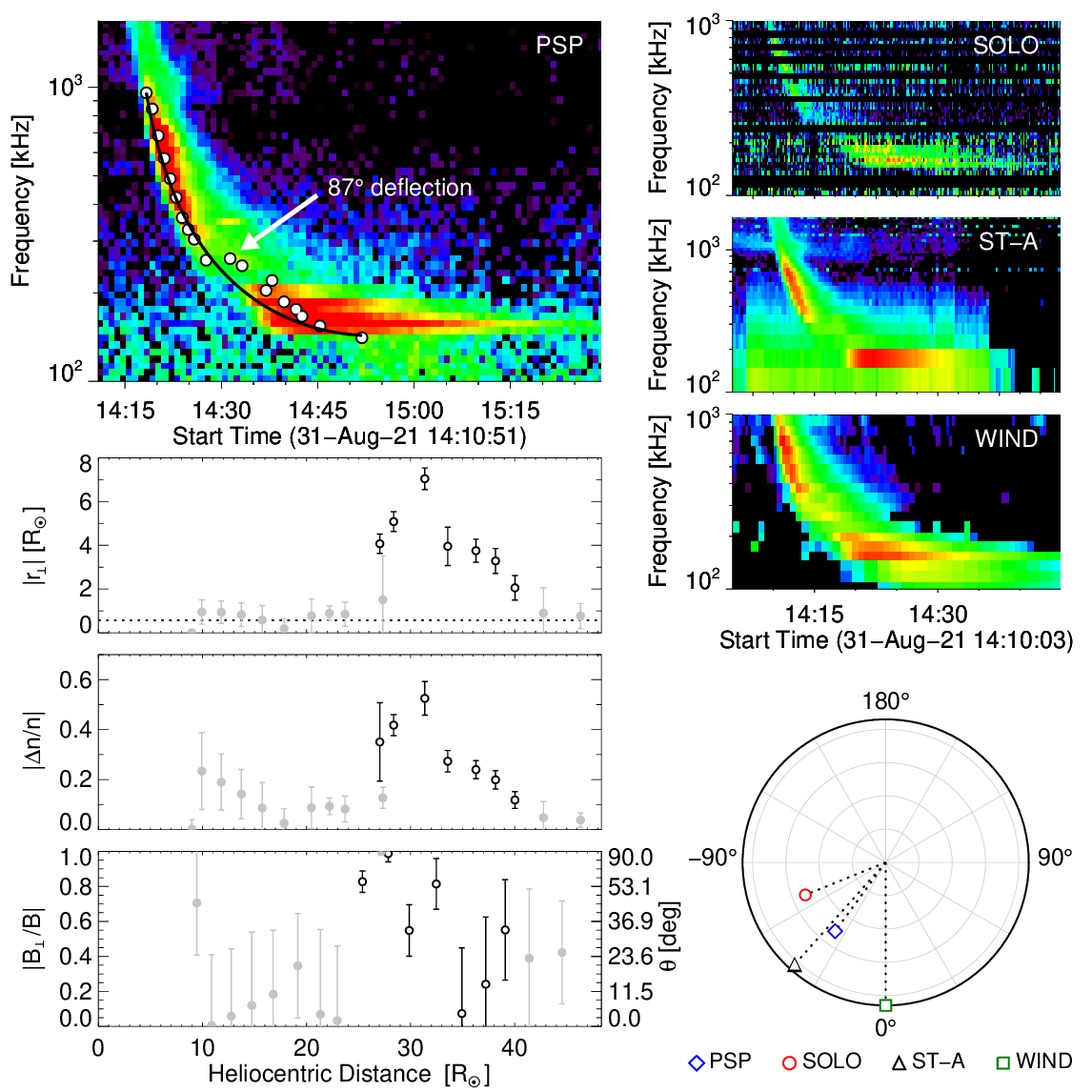}
        \caption{As in Figure \ref{fig:typeIII_features1} for a burst observed by PSP near 14:15 on 2021-Aug-31.}
        \label{fig:typeIII_features2}
    \end{figure}

\begin{figure}[htb!]
        \centering
        \includegraphics[width=0.8\textwidth]{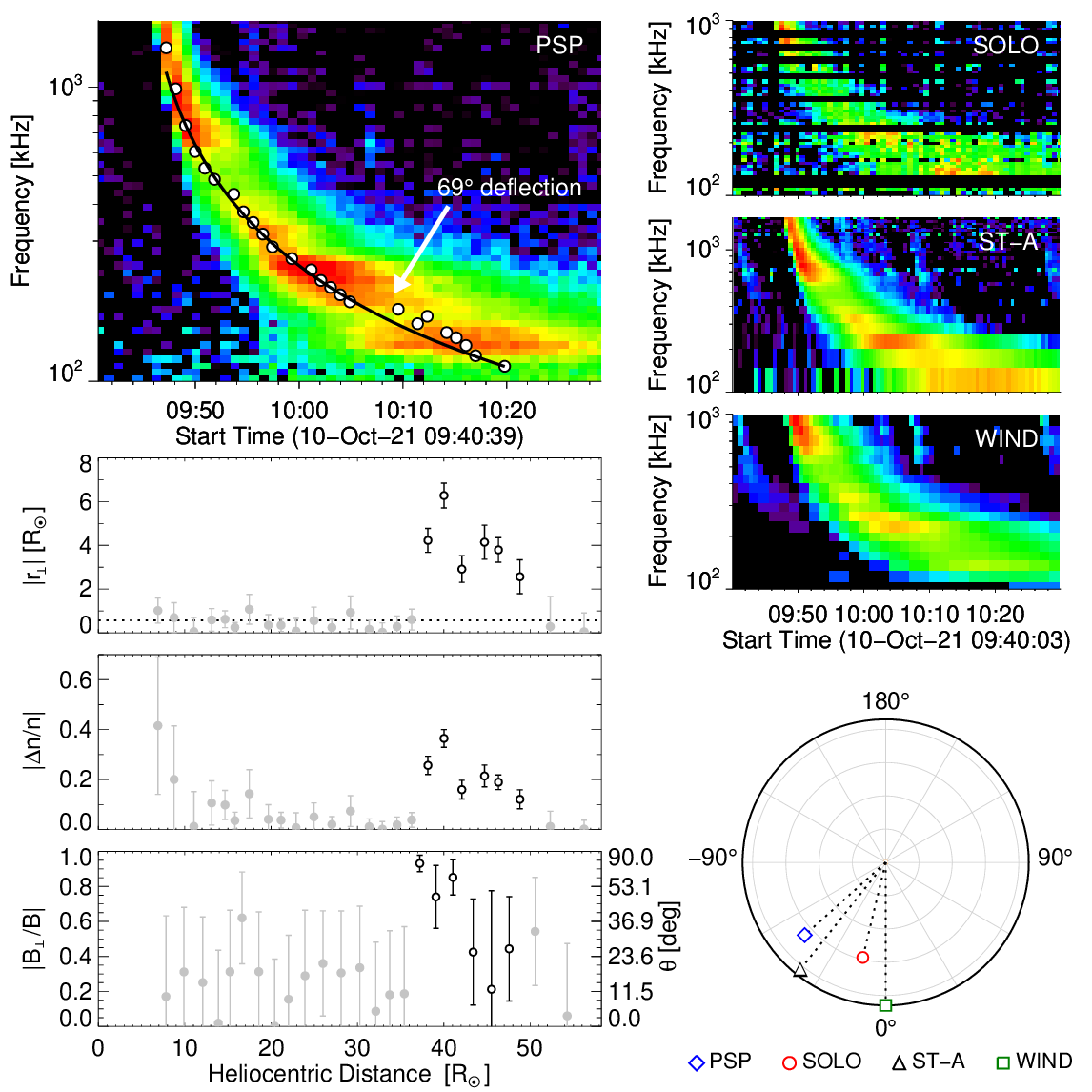}
        \caption{As in Figure \ref{fig:typeIII_features1} for a burst observed by PSP near 09:50 on 2021-Oct-10.}
        \label{fig:typeIII_features3}
    \end{figure}

\begin{figure}[htb!]
        \centering
        \includegraphics[width=0.8\textwidth]{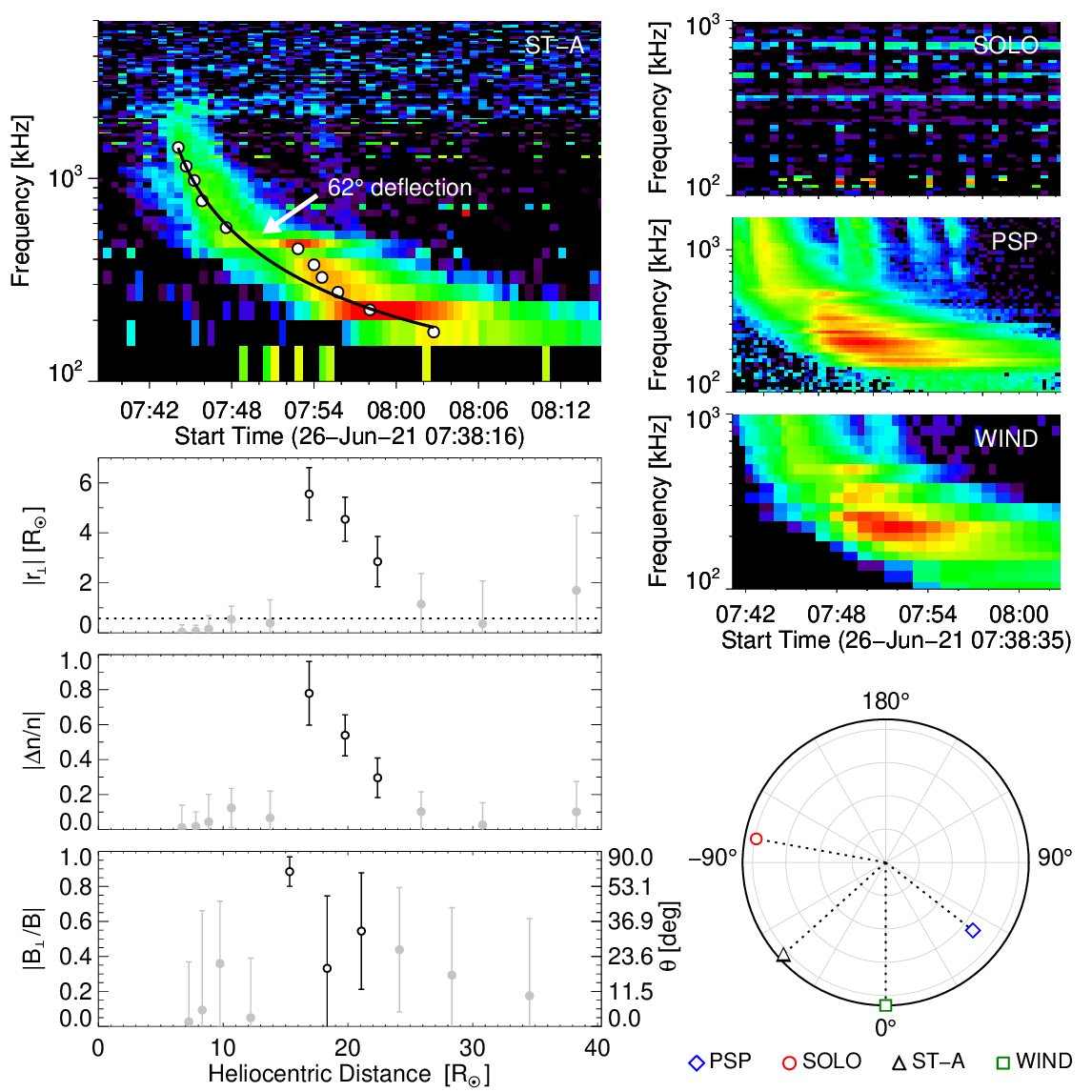}
        \caption{As in Figure \ref{fig:typeIII_features1} for a burst observed by ST-A near 07:44 on 2021-Jun-21. The other spacecrafts (right panel) show overlapping complex bursts making interpretation ambiguous.}
        \label{fig:typeIII_features4}
    \end{figure}

    Figures \ref{fig:typeIII_features2}--\ref{fig:typeIII_features4} present additional type III bursts with features that could be attributed to magnetic field disturbances, as discussed in section \ref{sec:obs_features}.

    
\clearpage
\bibliographystyle{aasjournal}
\bibliography{refs}

\end{document}